\documentclass[aps,prd,showpacs,eqsecnum,twocolumn,superscriptaddress,nofootinbib]
{revtex4-1}
\usepackage{amsmath,amssymb,graphicx,color}

\begin{document}

\title{Modeling GW170817 based on numerical relativity and its implications}

\author{Masaru Shibata}\affiliation{Center for Gravitational Physics,
  Yukawa Institute for Theoretical Physics, 
Kyoto University, Kyoto, 606-8502, Japan} 

\author{Sho Fujibayashi}\affiliation{Center for Gravitational Physics,
  Yukawa Institute for Theoretical Physics, 
Kyoto University, Kyoto, 606-8502, Japan} 

\author{Kenta Hotokezaka}\affiliation{School of Natural Sciences,
  Institute for Advanced Study, Princeton, NJ, 08540, USA}
\affiliation{Center for Gravitational Physics, Yukawa Institute for
  Theoretical Physics, Kyoto University, Kyoto, 606-8502, Japan}

\author{Kenta Kiuchi}\affiliation{Center for Gravitational Physics,
  Yukawa Institute for Theoretical Physics, 
Kyoto University, Kyoto, 606-8502, Japan} 

\author{Koutarou Kyutoku} \affiliation{
Theory Center, Institute of Particle and Nuclear Studies, KEK,
Tsukuba 305-0801, Japan\\
Department of Particle and Nuclear Physics, the Graduate University
for Advanced Studies (Sokendai), Tsukuba 305-0801, Japan\\
Interdisciplinary Theoretical Science (iTHES) Research Group, RIKEN,
Wako, Saitama 351-0198, Japan
}\affiliation{Center for Gravitational Physics,
  Yukawa Institute for Theoretical Physics, 
Kyoto University, Kyoto, 606-8502, Japan} 

\author{Yuichiro Sekiguchi} \affiliation{Department of Physics, Toho
  University, Funabashi, Chiba 274-8510, Japan}\affiliation{Center for
  Gravitational Physics, Yukawa Institute for Theoretical Physics,
  Kyoto University, Kyoto, 606-8502, Japan}

\author{Masaomi Tanaka} \affiliation{National Astronomical Observatory
  of Japan, Mitaka, Tokyo 181-8588, Japan}

\date{\today}

\newcommand{\beq}{\begin{equation}}
\newcommand{\eeq}{\end{equation}}
\newcommand{\beqn}{\begin{eqnarray}}
\newcommand{\eeqn}{\end{eqnarray}}
\newcommand{\pa}{\partial}
\newcommand{\vp}{\varphi}
\newcommand{\varep}{\varepsilon}
\newcommand{\ep}{\epsilon}
\begin{abstract}

Gravitational-wave observation together with a large number of
electromagnetic observations shows that the source of the latest
gravitational-wave event, GW170817, detected primarily by advanced LIGO,
is the merger of a binary neutron star. We attempt to interpret this
observational event based on our results of numerical-relativity
simulations performed so far paying particular attention to the optical
and infra-red observations.  We finally reach a conclusion that this
event is described consistently by the presence of a long-lived
hypermassive or supramassive neutron star as the merger remnant, because
(i) significant contamination by lanthanide elements along our line of
sight to this source can be avoided by the strong neutrino irradiation
from it and (ii) it could play a crucial role to produce an ejecta
component of appreciable mass with fast motion in the post-merger phase.
We also point out that (I) the neutron-star equation of state has to be
sufficiently stiff (i.e., the maximum mass of cold spherical neutron
stars, $M_{\rm max}$, has to be appreciably higher than $2M_\odot$) in
order that a long-lived massive neutron star can be formed as the merger
remnant for the binary systems of GW170817, for which the initial total
mass is $\agt 2.73M_\odot$ and (II) no detection of relativistic optical
counterpart suggests a not-extremely high value of $M_{\rm max}$
approximately as 2.15--$2.25M_\odot$.

\end{abstract}

\pacs{04.25.D-, 04.30.-w, 04.40.Dg}

\maketitle

\section{Introduction}\label{sec1}

On August 17, 2017, two advanced LIGO detectors (with an important
assistance by advanced VIRGO) succeeded in the first direct detection of
gravitational waves from an inspiraling binary system of two neutron
stars, which is referred to as GW170817~\cite{LIGO817}.  The data
analysis for this gravitational-wave event derives that the chirp mass,
defined by ${\cal M}:=(m_1 m_2)^{3/5}/(m_1+m_2)^{1/5}$ (where $m_1$ and
$m_2 (\leq m_1)$ denote each mass of the binary), is $\approx
1.188^{+0.004}_{-0.002}M_\odot$ for the 90\% credible interval.  This
implies that the total mass $m:=m_1 + m_2 = 2.729
(\eta/0.25)^{-3/5}M_\odot \geq 2.729M_\odot$.  Here, $\eta$ denotes the
symmetric mass ratio defined by $\eta:=m_1 m_2/m^2 (\leq 0.25)$.  The
mass ratio of the binary is not well constrained as 0.7--1.0 within the
90\% credible interval under the assumption that the dimensionless spin
of each neutron star is reasonably small ($\leq 0.05$).  However, the
values of $\eta$ for this mass-ratio range are between 0.242 and
0.250. This implies that the total mass is well constrained in the range
between $\approx 2.73M_\odot$ and $\approx 2.78M_\odot$ for the 90\%
credible interval.

The luminosity distance to the source from the earth is approximately
$D=40^{+8}_{-14}$\,Mpc~\cite{LIGO817}, and follow-up optical
observations (e.g., Ref.~\cite{LIGO817em} for a summary) found a
counterpart of this event and identified a S0 galaxy, NGC\,4993, as
the host galaxy. Since the sky location is accurately determined and
the total signal-to-noise ratio (SNR) of the gravitational-wave signal
is as high as 32.4~\cite{LIGO817}, the inclination angle of the binary
orbital axis with respect to our line of sight is constrained to be
$\iota \alt 28^{\circ}$~\cite{LIGO817}, and the effective distance to
the source (after taking into account the orbital inclination and sky
location with respect to the detector's orbital planes) is estimated
to be $D_{\rm eff}\approx 57$\,Mpc~\cite{LIGO817}.

A large number of observations in the optical and infra-red (IR) bands
have been also carried out following the gravitational-wave detection
(e.g.,
Refs.~\cite{EM0,EM1,EM2,EM3,EMSSS,EM4,EM5,EM6,EM7,EM8,EM9,EMNatAst}). These
observations show that the emission properties are largely consistent
with the macronova/kilonova model~\cite{Li,Metzger2010}, suggesting that
high-velocity, neutron-rich matter of mass 0.01--$0.1M_\odot$ ejected
from the neutron-star mergers radioactively shines through the
$r$-process nucleosynthesis~\cite{LS74,DLPS89} in the optical--IR bands
for 0.5--20 days after the merger, and that the spectrum is broadly
consistent with the quasi-thermal spectrum with significant reddening.
However, (i) the peak time of the light curve is earlier than the
expectation from a macronova/kilonova model in which heavy $r$-process
elements are appreciably synthesized and the typical value of the
opacity is expected to be $\kappa \approx 10\,{\rm cm^2/g}$ due to the
appreciable presence of lanthanide
elements~\cite{opacity,BK2013,TH2013,Tanaka17}, and (ii) the peak
luminosity is higher than what the typical scenarios have predicted for
the dynamical ejecta of binary neutron star mergers. A naive
interpretation for these observational results is that a fraction of the
ejecta is composed of lanthanide-poor material and the total ejecta mass
would be $\sim
0.025$--$0.05M_\odot$~\cite{EM0,EM1,EM2,EM3,EMSSS,EM4,EM5,EM6,EM7,EM8,EM9},
which is somewhat larger than the typical dynamical ejecta mass of $\sim
0.001$--$0.01M_\odot$ obtained by numerical-relativity simulations for
binary neutron star mergers.

In this paper, we attempt to interpret the results of the
electromagnetic observations for the optical--IR bands in terms of the
results of our wide variety of numerical-relativity simulations
performed so far. Numerical-relativity simulations for the merger of
binary neutron stars have been performed in our group since
1999~\cite{S1999,SU2000}, and now, detailed modeling for this
phenomenon is feasible as we describe in this paper.  We thus use the
latest numerical-relativity results for the interpretation of the
GW170817 event.

Fermi GBM and INTEGRAL reported a possible detection of an extremely
weak short gamma-ray burst (GRB) of duration 2\,s and the (isotropic)
luminosity $\sim 10^{47}$\,erg/s at $\sim 1.7$\,s after the trigger of
the GW170817 event~\cite{GRBGW,GBM,Integral}.  Since the binary
orbital axis with respect to our line of sight is likely to be mildly
misaligned with $\iota \alt 28^{\circ}$~\cite{LIGO817}, this
observation suggests a detection of an off-axis gamma-ray burst
emission or cocoon emission arising from an ultra-relativistic jet
launched at the merger \cite{lazzati,gottlieb}.  However, the
production of such an ultra-relativistic jet in numerical-relativity
simulations is beyond the scope of our paper.  Thus, we focus on
interpreting the optical and IR data in the following.

The paper is organized as follows. In Sec.~\ref{sec2}, we summarize
possible scenarios for the merger processes of binary neutron stars
with the total mass $m=2.7$--$2.8M_\odot$ and mass ratio
$q=m_2/m_1=0.8$--1.0 in the current constraint for the neutron-star
equation of state (EOS).  We note that there are seven Galactic
compact binary neutron stars observed to date~\cite{tauris17}. The
mass ratio of these binaries is in the range between $\approx 0.75$
and $\approx 1$, and the dimensionless spin of neutron stars, for
which the spin period is measured, is smaller than 0.03. Thus, in this
paper we do not consider extreme cases with small mass ratio like
$\leq 0.7$ or with a rapidly spinning neutron star.  We then discuss
in Sec.~\ref{sec3} what are special features for the observations of
GW170817, and draw a conclusion that the key point for describing this
event is the presence of a long-lived massive neutron star (either a
hypermassive or supramassive neutron star: see
Refs.~\cite{BSS00,NR2016} for their definition) as the remnant of the
binary neutron star merger, because significant contamination by
lanthanide elements along our line of sight to this source can be
avoided by the strong neutrino irradiation from it and also because it
could play a crucial role to produce an ejecta component of
appreciable mass with the fast motion of the velocity $\sim
0.1$--$0.2c$.  We also point out that if the long-lived massive
neutron star is indeed formed, this implies that the EOS has to be
stiff enough (i.e., the maximum mass of cold spherical neutron stars
has to be high enough) to escape the formation of a black hole in a
short time scale after the merger for the system of total mass $m \agt
2.73M_\odot$. Section~\ref{sec4} is devoted to discussing implications
of GW170817 and perspectives for the future observation. We then
summarize this paper in Sec.~\ref{sec5}.  Throughout this paper, $c$
denotes the speed of light.

\begin{table*}[t]
\centering
\caption{\label{tab2} Equations of state employed, the maximum mass
  for cold spherical neutron stars, $M_{\rm max}$, in units of the
  solar mass, the radius, $R_{M}$, and the dimensionless tidal
  deformability $\Lambda_{M}$ of spherical neutron stars of
  gravitational mass $M=1.20$, 1.30, 1.40, and $1.50M_\odot$.  $R_M$
  is listed in units of km. The last five data show the binary tidal
  deformability for $\eta=0.250$, 0.248, 0.246, 0.244, and 0.242 with
  ${\cal M}=1.19M_\odot$.  }
\begin{tabular}{cccccccccccc}
\hline\hline
~EOS~ & ~$M_{\rm max}$~
& ~$R_{1.20}$~ & ~$R_{1.30}$~ &  ~$R_{1.40}$~ & ~$R_{1.50}$~ 
& ~$\Lambda_{1.20}$~ &~$\Lambda_{1.30}$~ & 
~$\Lambda_{1.40}$~ &~$\Lambda_{1.50}$~ &
~$\Lambda$~ 
\\ \hline
SFHo & 2.06 
& 11.96 & 11.93 & 11.88 & 11.83
& ~864~ & ~533~ &  ~332~ & ~208~ 
& ~388, 387, 387, 386, 385~ \\
DD2  &2.42  
& 13.14 & 13.18 & 13.21 & 13.24
& ~1622~ & ~1053~ & ~696~ & ~467~ 
& ~797, 788, 780, 772, 764~ \\
\hline\hline
\end{tabular}
\end{table*}

\begin{table*}[t]
\centering
\caption{\label{tab1} Merger remnants and properties of dynamical
  ejecta for two finite-temperature neutron-star EOSs, SFHo and DD2
  and for the cases with different mass. The results of our radiation
  hydrodynamics simulations, in which both the neutrino heating and
  cooling are taken into account, are listed.  The quantities for the
  remnants are determined at $\approx 30$\,ms after the onset of
  merger.  HMNS, BH, and MNS denote hypermassive neutron star, black
  hole, and massive (hypermassive or supramassive) neutron star,
  respectively. The torus mass for the DD2 EOS is determined from the
  mass located outside the central region of MNS with density $\rho
  \leq 10^{13}\,{\rm g/cm^3}$ (left) and $\leq 10^{12}\,{\rm g/cm^3}$ (right).  The
  values of mass are shown in units of $M_\odot$. The BH spin means
  the dimensionless spin of the remnant black hole. $\bar Y_e$ and
  $\bar v_{\rm ej}$ are the average value of the electron fraction,
  $Y_e$, and average velocity of the dynamical ejecta,
  respectively. We note that $Y_e$ is broadly distributed between
  $\sim 0.05$ and $\sim 0.5$, irrespective of the models~(see
  Refs.~\cite{sekig15,sekig16}).  The ejecta mass has uncertainty by a
  factor of $\sim 2$ (see Appendix A for a discussion).}
\begin{tabular}{cccccccccc}
\hline\hline
~~EOS~~ & $m_1$ \& $m_2$ & $m_2/m_1$& Remnant & ~BH mass~ & ~BH spin~ & 
~Torus mass~ & ~~~$M_{\rm ej}$~~~ & ~~~ $\bar Y_e$ ~~~ & ~~~$\bar v_{\rm ej}/c$
~~~  \\ \hline
SFHo & 1.35, 1.35 & 1.00 & HMNS $\rightarrow$ BH & 2.59 & 0.69 & 0.05 & 0.011 & 0.31 & 0.22 \\
SFHo & 1.37, 1.33 & 0.97 & HMNS $\rightarrow$ BH & 2.59 & 0.70 & 0.06 & 0.008 & 0.30 & 0.21 \\
SFHo & 1.40, 1.30 & 0.93 & HMNS $\rightarrow$ BH & 2.58 & 0.67 & 0.09 & 0.006 & 0.27 & 0.20 \\
SFHo & 1.45, 1.25 & 0.86 & HMNS $\rightarrow$ BH & 2.58 & 0.69 & 0.12 & 0.011 & 0.18 & 0.24 \\
SFHo & 1.55, 1.25 & 0.81 & HMNS $\rightarrow$ BH & 2.69 & 0.76 & 0.07 & 0.016 & 0.13 & 0.25 \\
SFHo & 1.65, 1.25 & 0.76 & BH                    & 2.76 & 0.77 & 0.09 & 0.007 & 0.16 & 0.23 \\
DD2 & 1.35, 1.35 & 1.00 & MNS  & --- & --- & 0.23, 0.13 & 0.002 & 0.30 & 0.16 \\
DD2 & 1.40, 1.30 & 0.93 & MNS  & --- & --- & 0.23, 0.11 & 0.003 & 0.26 & 0.18 \\
DD2 & 1.45, 1.25 & 0.86 & MNS  & --- & --- & 0.30, 0.19 & 0.005 & 0.20 & 0.19 \\
DD2 & 1.40, 1.40 & 1.00 & MNS  & --- & --- & 0.17, 0.09 & 0.002 & 0.31 & 0.16 \\
\hline\hline
\end{tabular}
\end{table*}

\section{Summary of numerical-relativity results}\label{sec2}

In this section, we summarize the possible merger and post-merger
processes, in particular focusing on the merger remnant, ejecta mass,
and electron fraction of the ejecta.  The first one is closely related
to the central energy source, which determines the mechanisms of mass
ejection. The latter two are the key quantities for describing the
properties of the electromagnetic signals associated with the mass
ejection.  In the following subsections, we first summarize the merger
process of binary neutron stars and its dependence on the EOSs 
employed, total mass, and mass ratio of a binary focusing on the first
$\sim 30$\,ms after the onset of merger.  Then, we discuss the
possible long-term evolution processes of the merger remnants and the
associated mass ejection. The emphasis is on the following point: The
merger process, post-merger remnant evolution, mass ejection process,
and properties of the ejecta depend strongly on the neutron-star EOS,
in particular, on the maximum mass of cold spherical neutron stars,
$M_{\rm max}$.

\subsection{Dynamical merger process and dynamical mass ejection}\label{sec2.1}

\begin{figure*}[t]
\begin{center}
\includegraphics[width=110mm]{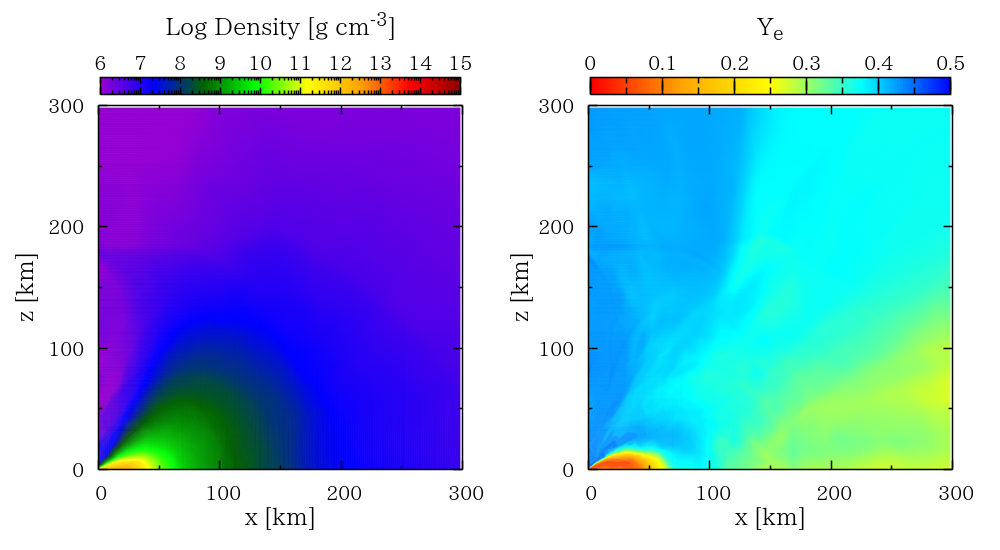} \\
\includegraphics[width=110mm]{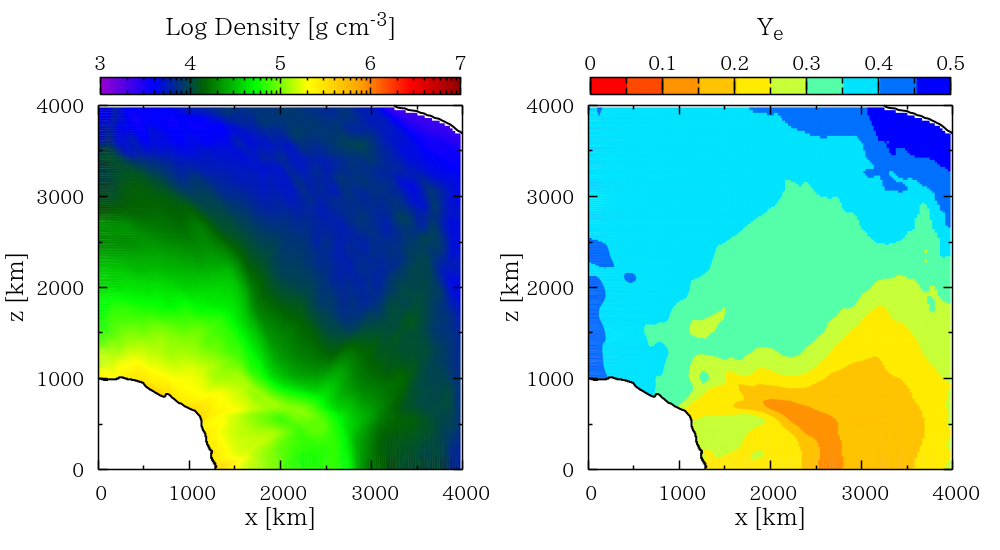}~~~~
\includegraphics[height=55mm]{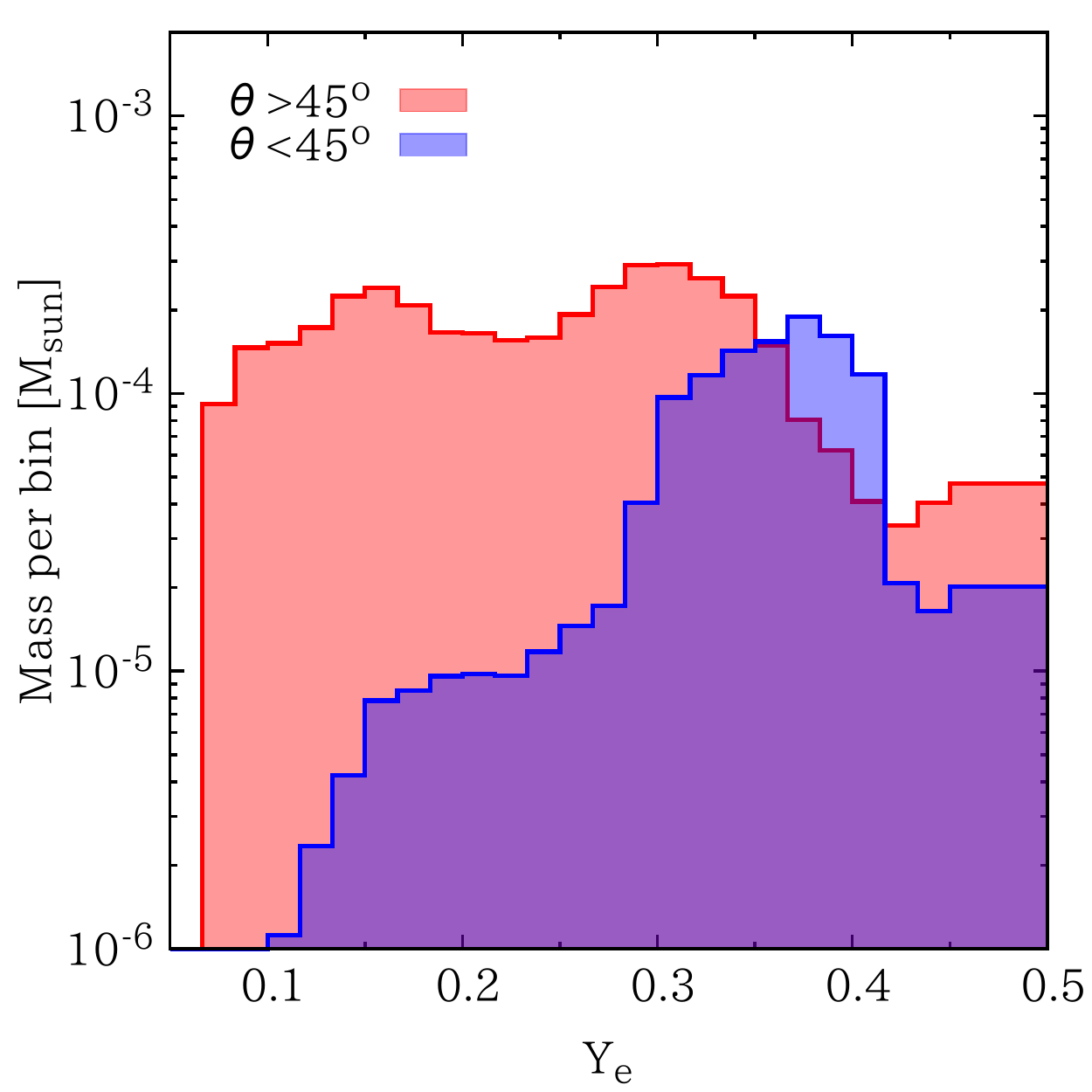}
\caption{Merger remnant for the SFHo EOS with $(m_1, m_2)=(1.4M_\odot,
  1.3M_\odot)$. The upper left and right panels show the profiles of the
  rest-mass density and electron fraction for the merger remnant, i.e.,
  a black hole surrounded by a torus, respectively.  The lower left and
  middle panels show the profiles of the rest-mass density and electron
  fraction for the dynamical ejecta component, respectively. The white
  region in these panels indicates that no ejecta component is present
  in the inner region.  These snapshots are generated at $\approx
  40$\,ms after the onset of merger. The lower right panel shows the
  mass histogram of the ejecta component as a function of $Y_e$ for the
  regions of $z > x~(\theta<45^\circ)$ and $z < x~(\theta > 45^\circ)$.
  \label{fig1}}
\end{center}
\end{figure*}

\begin{figure*}[t]
\begin{center}
\includegraphics[width=110mm]{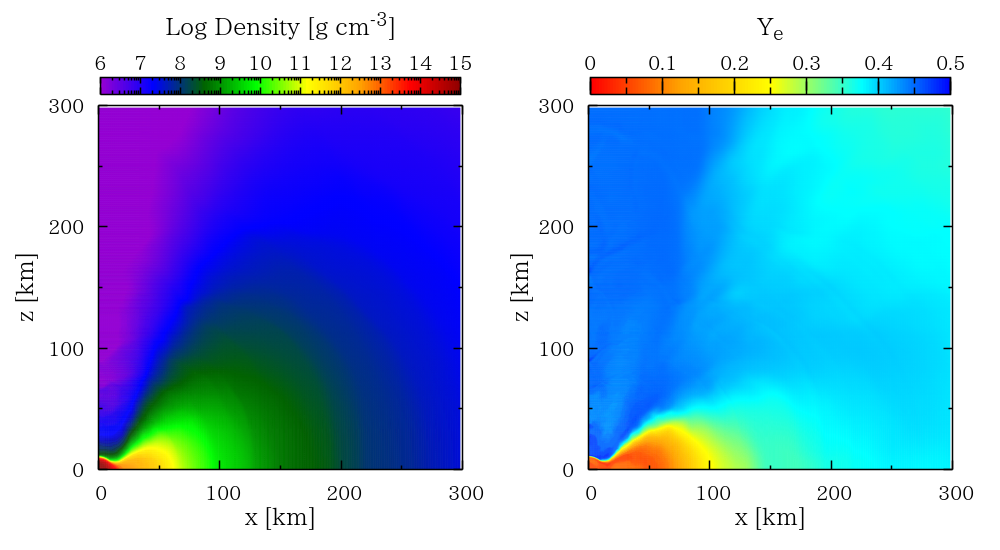} \\
\includegraphics[width=110mm]{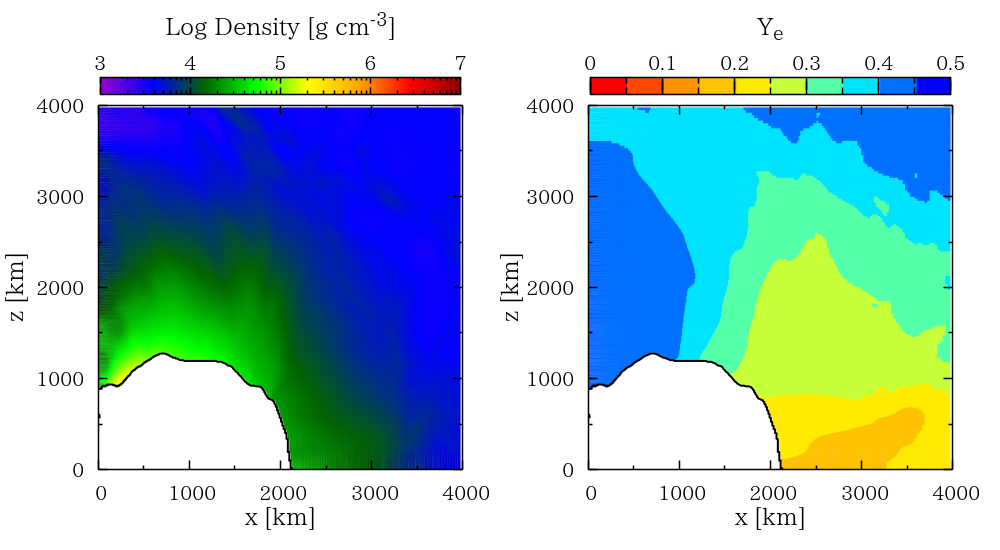}~~~~
\includegraphics[height=55mm]{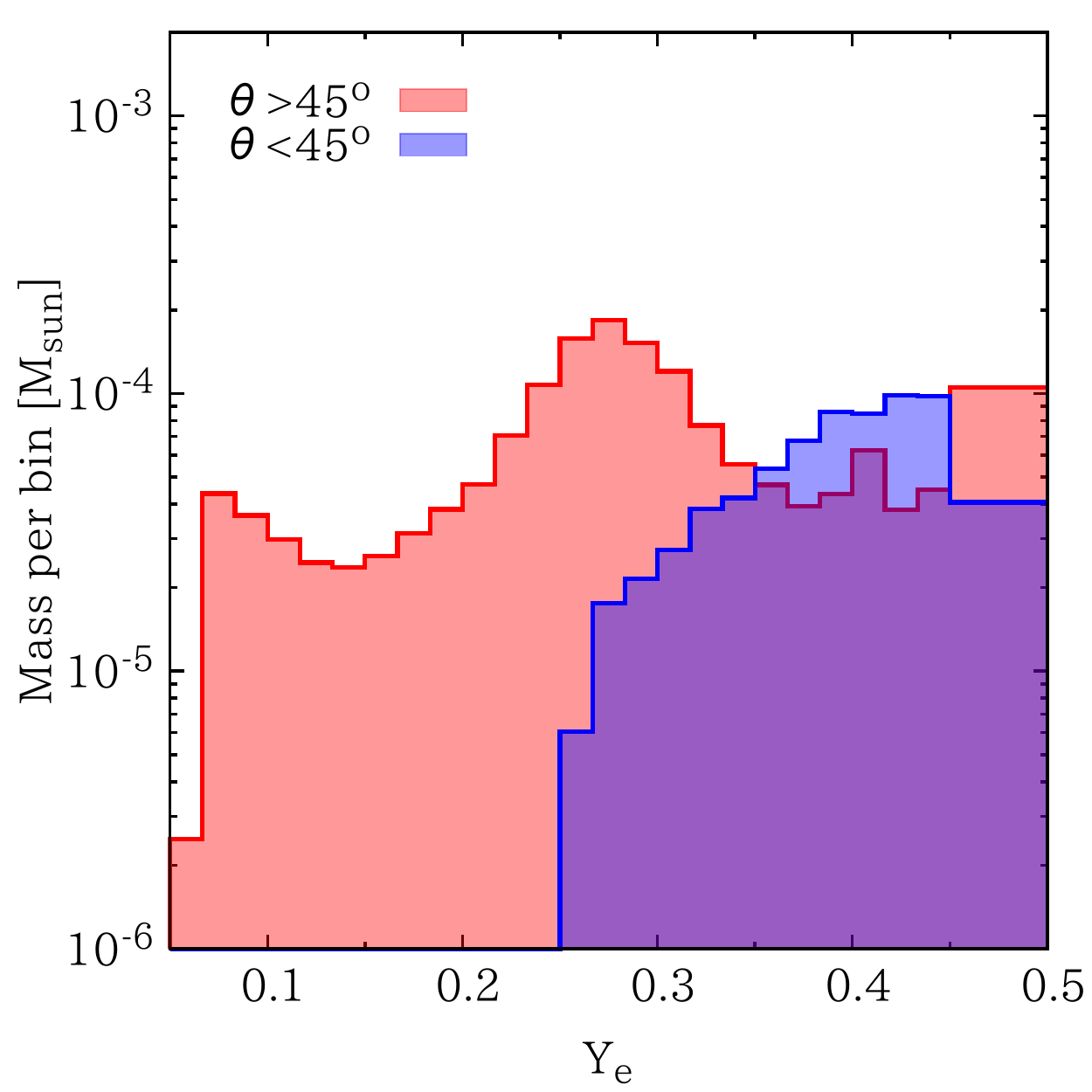}
\caption{The same as Fig.~\ref{fig1} but for the merger remnant for
  the DD2 EOS with $(m_1, m_2)=(1.4M_\odot, 1.3M_\odot)$.  The
  snapshots are generated at $\approx 80$\,ms after the onset of
  merger. For this case, a massive neutron star is located at the
  center (compare the upper panels of Figs.~\ref{fig1} and \ref{fig2}).
\label{fig2}}
\end{center}
\end{figure*}

A number of numerical-relativity simulations~(e.g.,
Refs.~\cite{Shibata050,Shibata05,Hotoke11,Hotoke13a,Hotoke13b,Hotoke13c,Hotoke13d,sekig15,sekig16})
have shown that the merger process and remnant object depend strongly
on the EOS of neutron stars, which is still poorly constrained.
However, because we approximately know the total mass of the binary
system for the event GW170817, we can discuss in detail the possible
merger process and remnant for a given hypothetical EOS. In the
following, we describe typical scenarios for the cases of soft and
stiff EOSs, for which the maximum mass for the cold spherical neutron
stars is $M_{\rm max} \alt 2.1M_\odot$ and $\agt 2.2M_\odot$,
respectively.  Popular soft and stiff EOSs, which are often employed in
the community of numerical relativity, is SFHo~\cite{SFHo} and DD2
EOSs~\cite{DD2}, for which $R \approx 11.9$\,km and 13.2\,km and
$M_{\rm max}=2.06M_\odot$ and $2.42M_\odot$, respectively (see
Table~\ref{tab2}).  Thus, we discuss the possible merger process and
remnants picking up numerical-relativity results for these two
representative EOSs. We note that for these two EOSs, the typical radius
and maximum mass are positively correlated. However, for some EOSs like
APR4~\cite{APR4}, the maximum mass could be $2.2M_\odot$ even for $R <
12$\,km. By contrast, for EOSs like H4~\cite{H4}, the maximum mass is
often only slightly larger than $2M_\odot$ while $R \sim 13.5$\,km.
In the discussion of this paper, the most important quantity is the
maximum mass, $M_{\rm max}$, and the typical neutron-star radius, $R$,
is not as important as the maximum mass.  We will touch on this point
in the final paragraphs of this subsection.

Table~\ref{tab1} summarizes the results of our numerical-relativity
simulations for the total mass $m=2.7$--$2.9M_\odot$ in the SFHo and
DD2 EOSs. Several numerical results in this table are taken from
Refs.~\cite{sekig15,sekig16}.  These simulations were performed taking
into account finite-temperature effects of nuclear-matter EOS,
neutrino cooling, and neutrino heating.  We note that including the
neutrino heating (irradiation) is quite important for predicting the
profile of the electron fraction, $Y_e$, for the merger remnants and
ejecta, and hence, in the following, we refer only to the 
numerical-relativity work in which this effect is taken into account.

In this subsection, we focus only on the dynamical ejecta that is
ejected in the first $\sim 30$\,ms after the onset of merger.  In
these simulations of the neutron-star mergers, no viscous nor
magnetohydrodynamics (MHD) effects are taken into account.  These are
likely to play key roles for the long-term evolution of the merger
remnants and could drive mass ejection in addition to the dynamical
ejecta. In Table~\ref{tab1}, such ejecta components are not
included. We will discuss the long-term evolution processes of the
merger remnant in the next subsection (see Sec.~\ref{sec2.2}).

\subsubsection{Summary of the dynamical ejecta}

Table~\ref{tab1} tells us the following facts: \\
\noindent
(i) For the SFHo models, hypermassive neutron stars are formed after
the merger temporarily for the total mass of $m=2.7$--$2.8M_\odot$,
but they subsequently collapse to a black hole surrounded by a torus
(see Fig.~\ref{fig1}). The lifetime of the hypermassive neutron stars
is $\alt 10$\,ms in this EOS model and is shorter for higher total
mass: For $m=2.8M_\odot$, it is $\sim 3$\,ms.  The mass and spin of
the remnant black holes are approximately $0.96m$ and $0.7$,
respectively, for the mass ratio $q \agt 0.8$.  The torus mass depends
strongly on $q$. However, for appreciably asymmetric binaries with $q
\alt 0.93$, the torus mass is likely to be larger than $\sim
0.1M_\odot$. For the high-mass torus, the typical electron fraction is
low, $Y_e\sim 0.1$ (see the upper panel of Fig.~\ref{fig1}), because
the density of the torus is high with the maximum value of $\sim
10^{12}\,{\rm g/cm^3}$, resulting in strong electron degeneracy and
strong neutronization.  It should be also noted that the outer region
with $x \alt 100$\,km is fairly neutron-rich with $Y_e=0.2$--0.4 for
this case. The property of low values of $Y_e$ is different from that
in the presence of a massive neutron star: see (ii) below.  The reason
for this is that the torus, which is the only source of the neutrino
emission in this remnant system, is a weak neutrino emitter (in the
absence of efficient viscous heating), and hence, the neutrino
irradiation is not efficient enough to increase the value of $Y_e$ for
the matter around the black hole (a possible effect of the viscous
heating will be discussed in Sec.~\ref{sec2.2}).

\noindent
(ii) For the DD2 models, a long-lived massive (perhaps supramassive)
neutron star surrounded by a torus is universally formed for
$m=2.7$--$2.8M_\odot$ (see Fig.~\ref{fig2}). The torus mass depends
weakly on the mass ratio and it is $\sim 0.2M_\odot$ for $q \sim 1$
and $\sim 0.3M_\odot$ for $q \sim 0.85$.  The electron fraction for
the high-density region of the torus is slightly higher than that in
the torus surrounding a black hole found in the SFHo models because
the matter in the torus experiences shock heating more efficiently
around the massive neutron stars.  A more remarkable fact is that in
the outer region of the torus at $x \agt 100$\,km, the electron
fraction is quite high as $Y_e \agt 0.25$ (see the upper panel of
Fig.~\ref{fig2}) because of the irradiation by neutrinos emitted from
the central massive neutron star and surrounding torus~(e.g.,
Refs.~\cite{Perego,sekig15,sekig16,Foucart2016}): In such an
environment, the neutrino capture processes, $n+\nu_e \rightarrow p +
e^{-}$ and $p+\bar \nu_e \rightarrow n + e^+$, take place quite
efficiently in the matter surrounding the massive neutron star and
torus, and by the balance of these reactions, the fraction of neutrons
and protons approaches an equilibrium value in which the value of
$Y_e$ is enhanced to be $\alt 0.5$ and approximately given by~(e.g.,
Ref.~\cite{QW96})
\begin{equation}
Y_{e, \rm{eq}} \sim \left[ 1 + \frac{L_{\bar{\nu}_{e}}}{L_{\nu_{e}}}\cdot
\frac{\langle \epsilon_{\bar{\nu}_{e}} \rangle - 2\Delta}
{\langle \epsilon_{\nu_{e}} \rangle + 2\Delta} \right]^{-1},
\end{equation}
where $\Delta = 1.293$\,MeV (the mass energy difference between
neutron and proton), $\langle \epsilon_{\nu_{e}} \rangle \approx
10$\,MeV and $\langle \epsilon_{\bar{\nu}_{e}} \rangle \approx
15$\,MeV denote the averaged neutrino energy of $\nu_e$ and
$\bar\nu_e$, respectively, and $L_{\nu_e} (\agt 10^{53}\,{\rm erg/s})$ and
$L_{\bar\nu_e}(\agt L_{\nu_e})$ denote the luminosity of $\nu_e$ and
$\bar\nu_e$, respectively.  Note that this enhancement of $Y_e$ is not
seen for the case that a black hole is formed soon after the merger
because of the absence of the strong neutrino source (compare the upper
panels of Figs.~\ref{fig1} and \ref{fig2}).

\noindent
(iii) For the SFHo models, the dynamical ejecta mass is $\sim
0.01M_\odot$ irrespective of the mass ratio (see
Table~\ref{tab1}). The electron fraction is distributed for a wide
range between 0.05 and 0.5~\cite{sekig15,sekig16} (see the lower
panels of Fig.~\ref{fig1}), also irrespective of the mass ratio. In
such ejecta, an appreciable fraction of lanthanide elements should be
synthesized~\cite{Oleg,Wanajo14,Kasen15,Tanaka17}, significantly
enhancing the opacity.  For this EOS, the dynamical mass ejection
occurs in a quasi-isotropic manner because not only the tidal effect
but also the shock heating play an important role for ejecting matter
(see the lower left panel of Fig.~\ref{fig1}).  Nevertheless, the
matter is ejected primarily along the binary orbital plane and ejecta
in the polar direction has a minor fraction, because of the strong
tidal effect during the merger and the presence of angular
momentum. The ejecta near the binary orbital plane is always neutron
rich with $Y_e \leq 0.25$ (see the lower panels of Fig.~\ref{fig1}),
although the ejecta in the polar region is less neutron-rich.

\noindent
(iv) For the DD2 models, the dynamical ejecta mass depends strongly on
the mass ratio: For $q=1$, it is $0.002M_\odot$, while for $q=0.86$ it
increases to $0.005M_\odot$.  The electron fraction is again widely
distributed for a range between 0.05 and 0.5~\cite{sekig15,sekig16}
(see also the lower panels of Fig.~\ref{fig2}) irrespective of the mass
ratio.  For this EOS, the matter is ejected primarily toward the
direction of the binary orbital plane, because the tidal effect during
the merger plays a dominant role for the mass ejection.  As in the
SFHo case, the ejecta in the binary orbital plane is neutron rich
 with $Y_e \leq 0.25$, in particular for the highly asymmetric-mass 
binaries. On the other hand, the ejecta in the polar region is less
neutron-rich with $Y_e \agt 0.25$ (see the lower-right panel of
Fig.~\ref{fig2}).

\noindent
(v) The average velocity of the dynamical ejecta is $0.15$--$0.25c$
depending on the EOS and mass ratio. For the SFHo case, the average
velocity is by 20--30\% larger than that for the DD2 case for a given
value of mass because, for this EOS, the neutron-star radius is small,
and hence, the shock heating effect during merger enhances the kinetic
energy of the ejecta.

In the above summary, the points worthy to note are as follows: (I) No
models predict the mass of the dynamical ejecta larger than
$0.02M_\odot$. This implies that if a luminous optical-IR counterpart
which requires the ejecta mass of $\geq 0.02M_\odot$ is discovered, we
have to consider ejecta components other than the dynamical ejecta.
(II) Irrespective of the EOS and binary mass ratio, the electron
fraction is widely distributed and the highly neutron-rich matter is
always present in the dynamical ejecta, in particular, near the binary
orbital plane. Only for the direction of the rotational axis of the
orbital motion ($\theta \alt 45^\circ$), the neutron richness is
suppressed resulting in $Y_e \agt 0.25$. (III) Material ejected toward
the polar direction ($\theta \alt 45^\circ$) is a minor component in
terms of the mass. Although the polar ejecta has a high value of
$Y_e=0.3$--0.4, this component does not contain a sufficient amount of
mass to produce a bright blue kilonova \cite{Kasen17}.

Nucleosynthesis studies (e.g. Refs.~\cite{Oleg,Wanajo14}) have shown
that the presence of neutron-rich ejecta with $Y_e \alt 0.25$ results in
producing a substantial fraction of lanthanide elements, and as a
result, the opacity of the ejecta is significantly enhanced to be
$\kappa \sim 10\,{\rm cm^2/g}$~\cite{opacity,BK2013,TH2013,Tanaka17}. As
we discuss in Sec.~\ref{sec3.1}, if high-mass and low-$Y_e$ ejecta are
present along our line of sight to the source, the peak time scale of
the optical light curve (in particular for optical to near IR light)
would be long $\agt 1$\,week (if the fraction of the lanthanide elements
is $\agt 10^{-4}M_\odot$~\cite{Kasen15}). This effect is often referred
to as the lanthanide curtain. However, the observations for GW170817
show that the optical light curve has a peak at $\alt
1$\,day~\cite{EM0,EM1,EM2,EM3,EMSSS,EM4,EM5,EM6,EM7,EM8,EM9}, suggesting
that $\kappa$ should be much smaller than $10\,{\rm cm^2/g}$ for the
early component of the optical-IR counterparts, and hence, the
contamination by lanthanide elements would be significantly suppressed
at least in the outer part of the ejecta along our line of sight.

\subsubsection{Remarks on neutron-star EOS}

In this section, we have focused on the models employing only two
representative EOSs. There are a wide variety of alternative
possibilities for the neutron-star EOS.  Here we point out that for some
extremely soft EOSs (which still can reproduce two-solar mass neutron
stars~\cite{demorest10}), a black hole is formed directly after the
onset of merger for the total mass $m \agt 2.7M_\odot$. Such models can
be constructed for EOSs in which the typical stellar radius is smaller
than 11\,km and the maximum mass for cold spherical neutron stars is
only slightly larger than $2M_\odot$. One such example is the so-called
B EOS which is one of piecewise polytropic EOSs composed of two
pieces~\cite{Read09b,kiuchi2017}. For this example, a black hole is
directly formed after the onset of merger for $m \geq 2.7M_\odot$.  For
the case that the mass asymmetry of the binary is not very large with
this type of EOS, any torus cannot be appreciably formed surrounding the
remnant black hole (e.g., Refs.~\cite{Shibata05,Kiuchi09}), and
moreover, the ejected mass cannot exceed $10^{-3}M_\odot$~(e.g.,
Ref.~\cite{Hotoke13a}).  With this model, the observed electromagnetic
counterparts of GW170817 cannot be described. As found for the model
with the SFHo EOS and with mass $(m_1, m_2)=(1.65M_\odot, 1.25M_\odot)$,
an appreciable mass ejection is still possible even for direct
black-hole formation in a case of high mass-asymmetry binaries. However,
in this case, the dynamical ejecta is extremely neutrino-rich so that
the short peak time of the electromagnetic counterparts for
GW170817~\cite{EM0,EM1,EM2,EM3,EMSSS,EM4,EM5,EM6,EM7,EM8,EM9} may not be
reproduced due to the lanthanide-curtain effect~\cite{Kasen15} (see the
discussion in Sec.~\ref{sec3.2}). These facts indicate that the
optical-IR counterparts of GW170817 can be used to rule out a group of
soft EOSs in which the stellar radius is small ($< 11$\,km) and the
value of $M_{\rm max}$ is not much larger than $2M_\odot$ (see also
Ref.~\cite{Bauswein} for an independent analysis).  We plan to further
explore this issue by numerical-relativity
simulations~\cite{Shibata2017}.

It should be also noted that even for an EOS in which $R > 13$\,km
(e.g., H4 EOS~\cite{H4}), the remnant massive neutron star could be
short-lived for $m \agt 2.75M_\odot$~(e.g.,
Ref.~\cite{Hotoke11,Hotoke13b}) if the value of $M_{\rm max}$ is $\sim
2M_\odot$. For this type of EOS, the post-merger process is likely to
be similar to that for the SFHo EOS. Thus, the value of $M_{\rm max}$
is a key quantity for the discussion of this paper.  We would like the
readers to keep this point in mind.

\subsection{Long-term evolution of merger remnants}\label{sec2.2}

\begin{figure*}[p]
\begin{center}
\includegraphics[width=110mm]{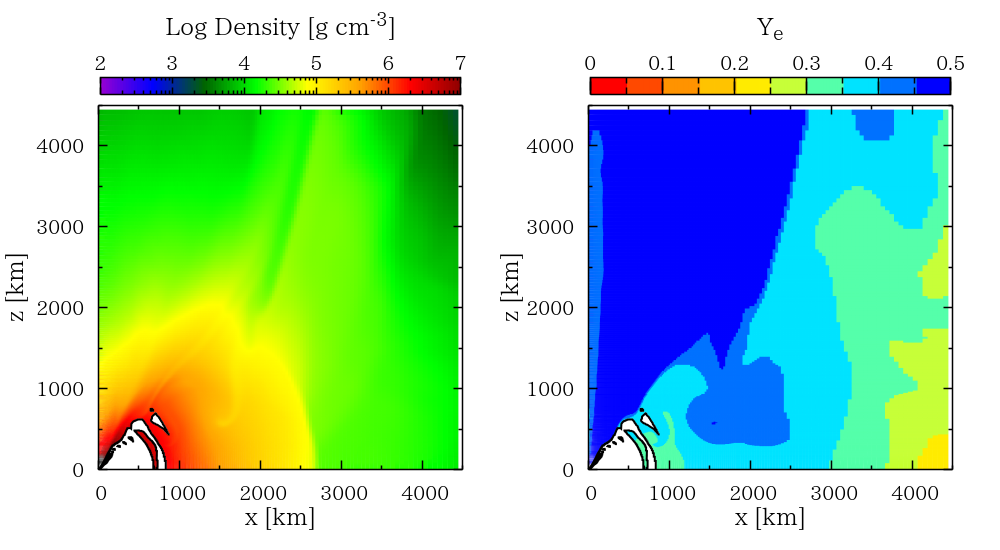}~~~~
\includegraphics[height=55mm]{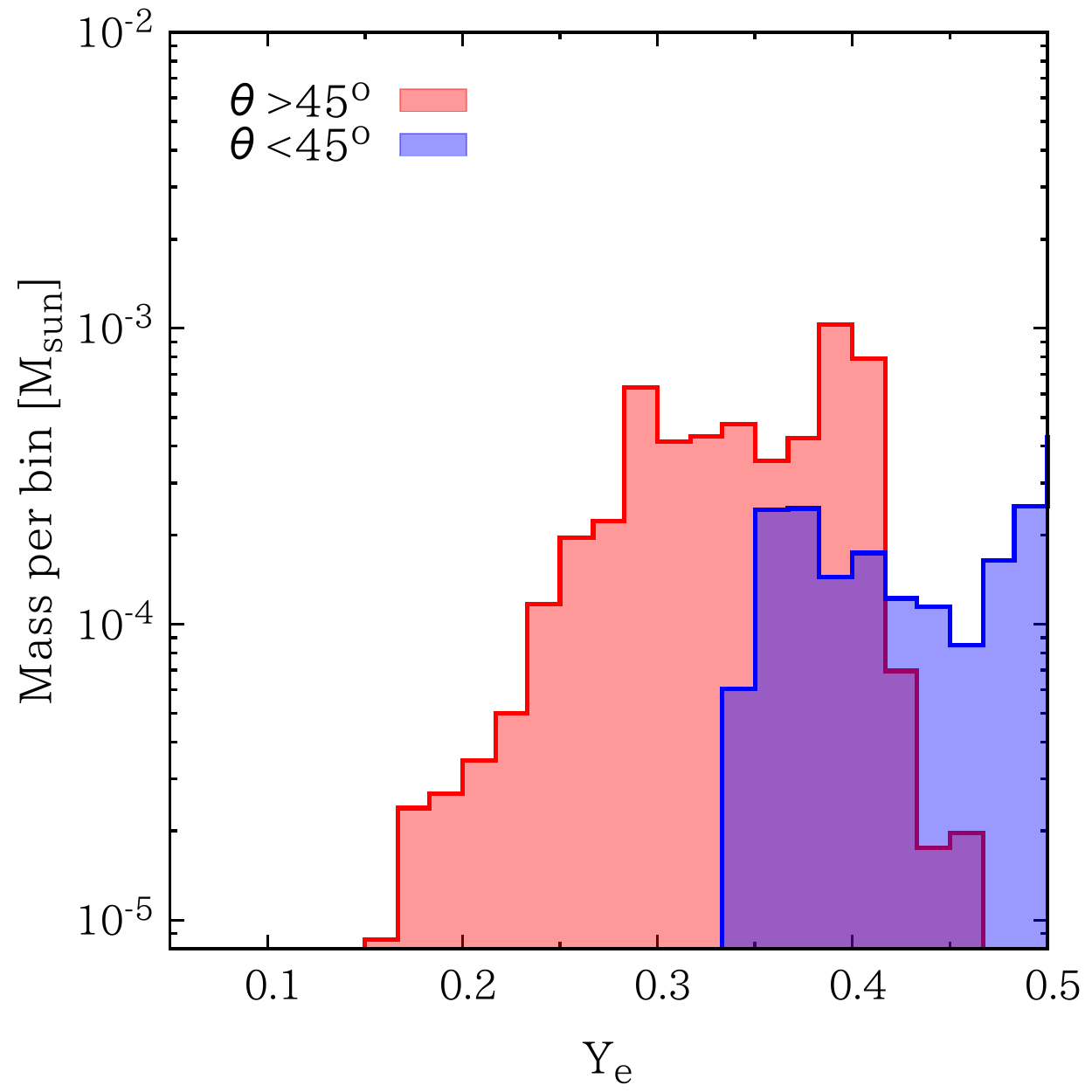} 
\\
\includegraphics[width=110mm]{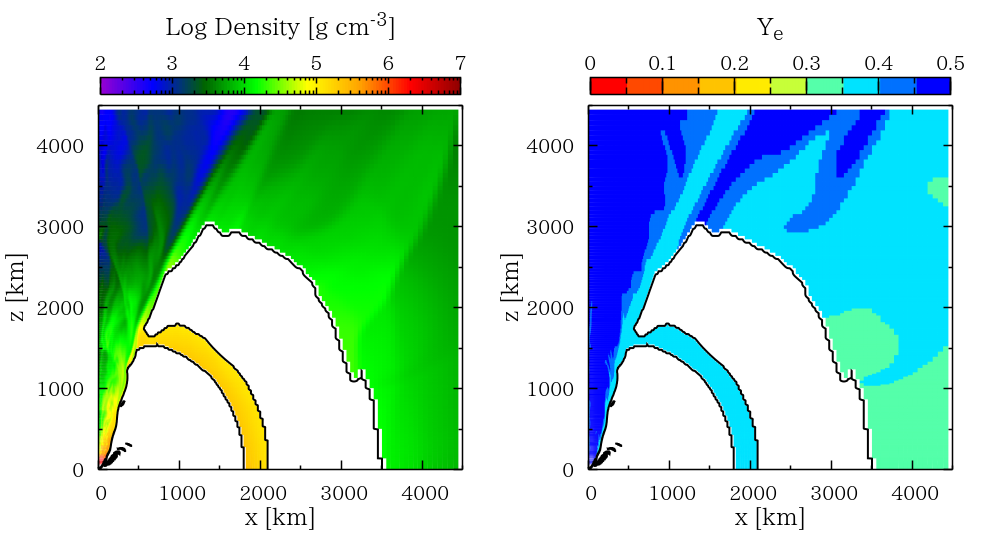}~~~~
\includegraphics[height=55mm]{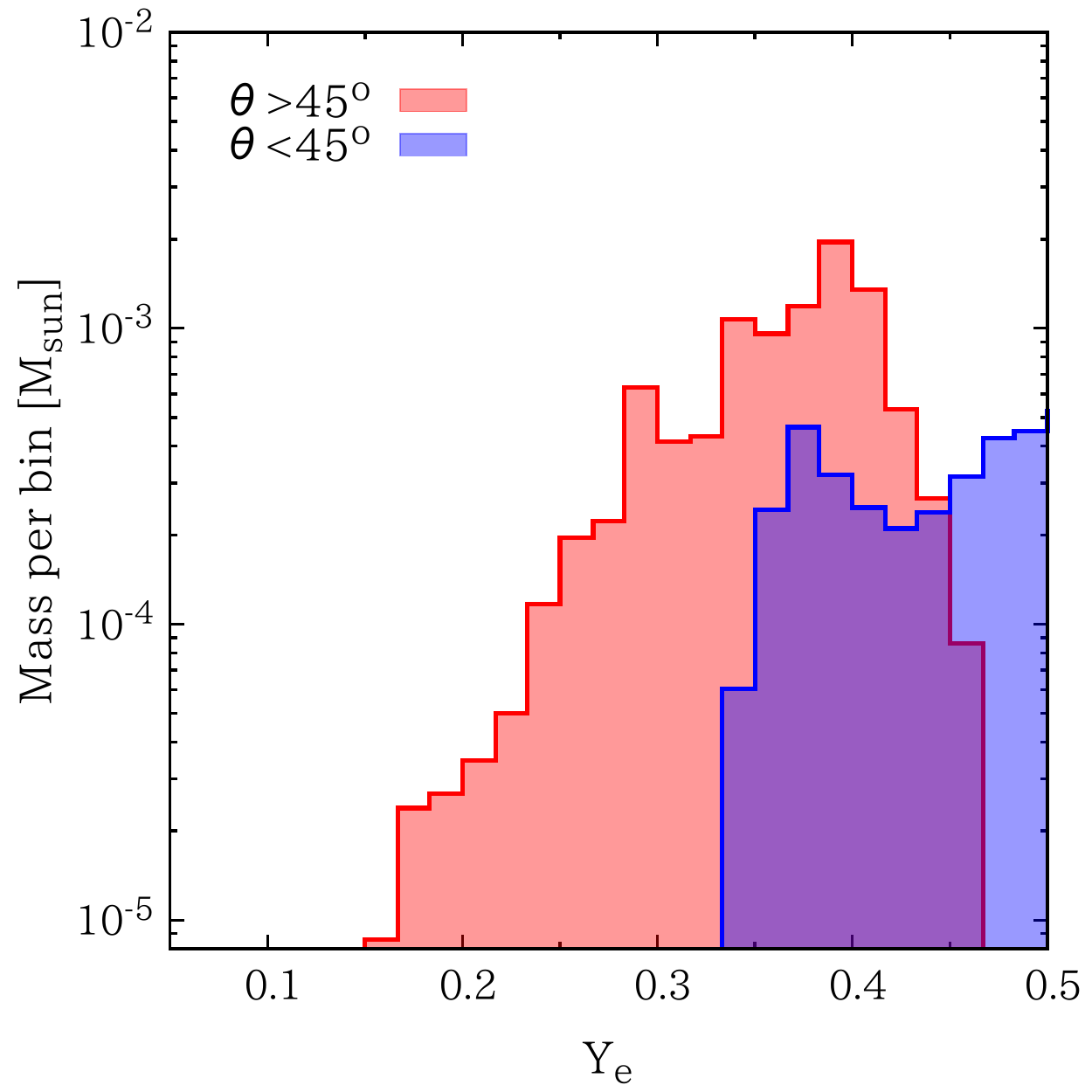} 
\\
\includegraphics[width=110mm]{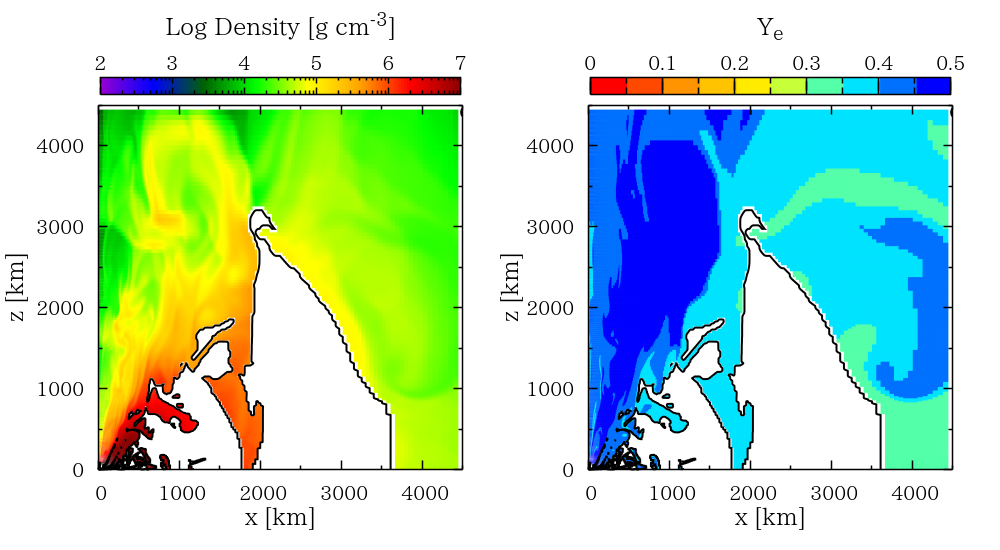}~~~~
\includegraphics[height=55mm]{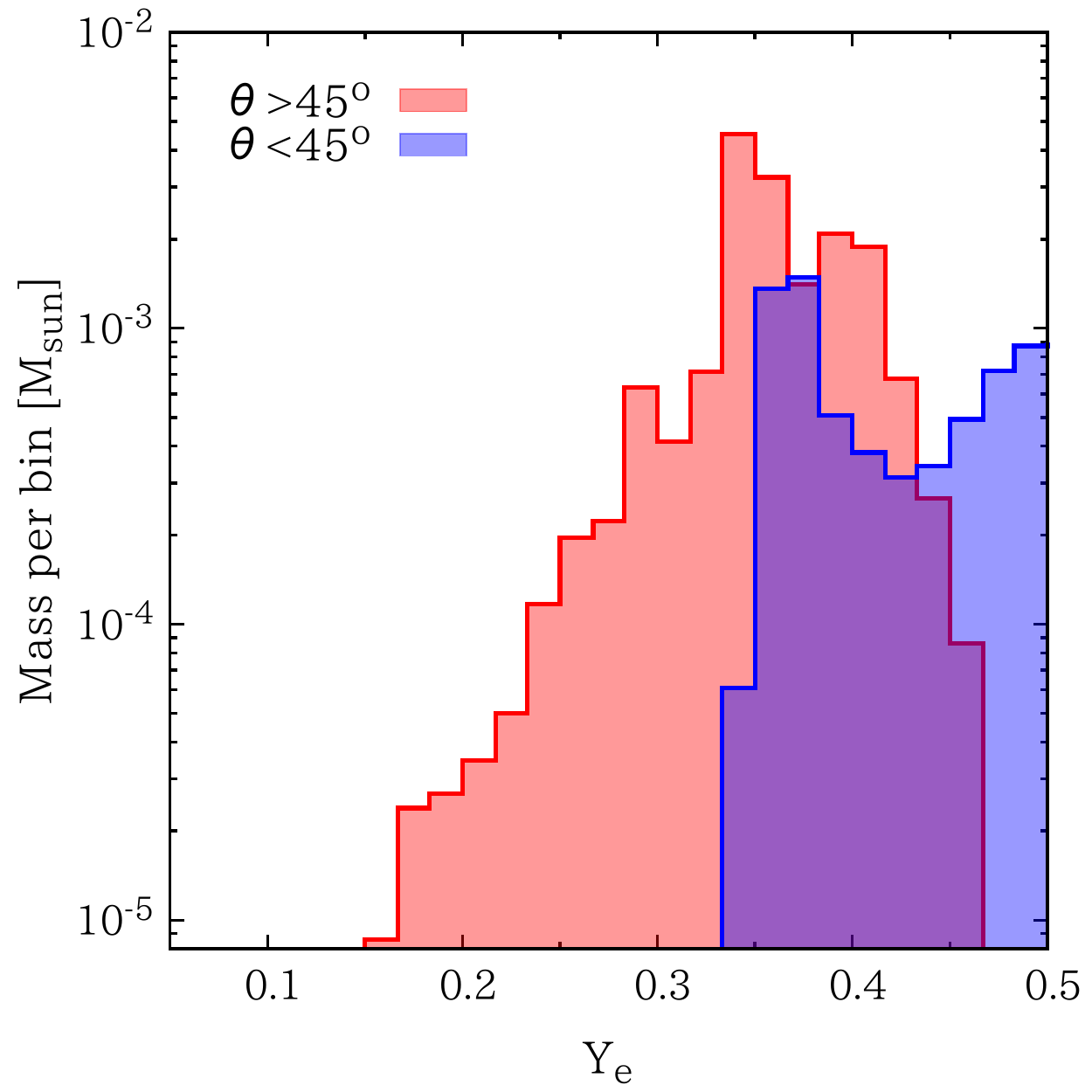} 
\caption{The upper panels: The profiles of the rest-mass density
  (left) and electron fraction (middle) for the early viscosity-driven
  ejecta component at $t=100$\,ms after the evolution of the massive
  neutron star-torus system. The white region indicates that no ejecta
  component is present in the inner region.  The right panel shows the
  mass histogram of the accumulated ejecta component as a function of
  $Y_e$ for the regions of $z > x~(\theta<45^\circ)$ and $z <
  x~(\theta > 45^\circ)$. The middle panels: The same as the upper
  panels but at $t=500$\,ms after the evolution of the system at which
  only the viscosity-driven ejecta from the torus with neutrino
  irradiation is dominant in the computational region.  The bottom
  panels: The same as the upper panels but at $t=1500$\,ms after the
  evolution of the system at which the late-time viscosity-driven
  ejecta from the torus is driven, increasing the ejecta of $0.3 \alt
  Y_e \alt 0.4$.  For all the panels, the results with $\alpha_{\rm
    vis}=0.04$ for the DD2 model are shown.
\label{fig6}}
\end{center}
\end{figure*}

A torus surrounding a remnant black hole or a massive neutron star
with a torus could be another source for the mass ejection because the
remnant torus and remnant massive neutron star are differentially
rotating, and hence, MHD/viscous effects induce angular momentum
transport and viscous heating, which could drive long-term mass
ejection, the so-called viscosity-driven mass
ejection~\cite{FM13,FM14,FM15,Just2015,SM17}. However, in the previous
simulations~\cite{FM13,FM15,Just2015,SM17}, basically, black holes
were considered as the central object. For some studies, massive
neutron stars were considered as the central object and the importance
of the neutrino irradiation was qualitatively pointed
out~\cite{FM14,Perego,Lippu17}. However, their treatment for remnant massive
neutron stars was rather artificial, and thus, their results are not
very conclusive.

Recently, we performed shear-viscous-radiation hydrodynamics
simulations in general relativity for a merger remnant~\cite{Fujiba17}
that is obtained from one of the numerical-relativity simulations
summarized in Sec.~\ref{sec2.1}.  Specifically, we performed
simulations for the remnant of the model with the DD2 EOS and
$m_1=m_2=1.35M_\odot$.  We evolved both the massive neutron star and
torus in a self-consistent manner. We first summarize the general
properties of mass ejection for this model.

\subsubsection{Evolution of massive neutron star-torus system}\label{sec3.2.1}

For our shear viscous hydrodynamics
simulations~\cite{Fujiba17,Shibata2017a}, we have to give the shear
viscous coefficient $\nu$. Using the $\alpha$-viscous
prescription~\cite{SS72,ST}, we set it as
\beq
\nu=\alpha_{\rm vis} H c_s,
\eeq
where $\alpha_{\rm vis}$ is the so-called dimensionless $\alpha$
parameter, $H$ is the maximum scale height of the systems, and $c_s$
is the sound speed. Since we are interested in the evolution of the
remnant massive neutron star and the torus surrounding it, we set
$H=10$\,km.  We employed $\alpha_{\rm vis}=0.01$, 0.02, and 0.04
following the finding in the latest high-resolution MHD simulations
for accretion disks~(e.g.,
Refs.~\cite{alphamodel,suzuki,local}). Our latest high-resolution
  MHD simulation~\cite{Kiuchi17} also shows that at least for an outer
  region of the remnant massive neutron star and torus, $\alpha_{\rm
    vis}$ is likely to be enhanced to $\sim 0.02$. We note that a
strong turbulent state is likely to be realized in the merger
remnants, because at the onset of merger, the Kelvin-Helmholtz
instability and subsequent quick winding of the magnetic fields could
significantly amplify the magnetic-field strength inside the remnant
massive neutron star to $\agt 10^{16}$\,G, and as a result, MHD
turbulence is likely to be induced for the
remnant~\cite{RR06,Kiuchi14,Kiuchi15,Kiuchi17}.

For the viscous evolution of a remnant massive neutron star surrounded
by a torus, there are two mechanisms for the mass
ejection~\cite{Fujiba17}.  In the short-term evolution with the
duration in a few tens of ms, the differential rotation of the remnant
massive neutron star becomes the engine for the mass ejection. In the
presence of viscosity, this differentially rotating state is changed
to a rigidly-rotating state in the viscous time scale of
\begin{align}
\frac{R_{\rm eq}^2}{\nu} \approx& 23\ {\rm ms}
\biggl(\frac{\alpha_{\rm vis}}{0.01}\biggr)^{-1} 
\biggl(\frac{c_s}{c/3}\biggr)^{-1} 
\biggl(\frac{R_{\rm eq}}{15\,{\rm km}}\biggr)^2 
\biggl(\frac{H}{10\,{\rm km}}\biggr)^{-1}, \nonumber \\
\label{eq:mnsvis}
\end{align}
where $R_{\rm eq}$ denotes the equatorial radius of the massive
neutron star.  During this transition, the density and pressure
profiles are changed on the short time scale, and associated with
this, strong density waves and resulting shock waves are generated and
propagate outward. Subsequently, spending a few tens of ms, these
shock waves sweep matter surrounding the central massive neutron star,
including the torus and atmosphere around it, and provide energy to
them. As a result, the matter in the outer region of the torus is
ejected in a quasi-isotropic manner~\cite{Fujiba17} ($\theta \agt
30^\circ$: see the upper panels of Fig.~\ref{fig6}).  We refer to this
mass ejection process as ``early viscosity-driven mass ejection''.  In
our numerical experiments, the mass of this ejection is $\approx
0.01(\alpha_{\rm vis}/0.02)M_\odot$ for $0.01 \leq \alpha_{\rm vis}
\leq 0.04$ for the case of a torus of mass $\sim 0.2M_\odot$.  This
implies that if a significantly strong turbulent state is achieved in
the remnant massive neutron star and the viscous parameter is
effectively enhanced to be $\agt 0.04$, significant mass $>
0.02M_\odot$ would be ejected.  Thus, the ejecta mass in this
mechanism could be $\sim 10$ times as large as the dynamical ejecta
mass for the DD2 model of nearly equal-mass
binaries~\cite{sekig15,sekig16}.

In this early viscosity-driven mass ejection, matter in the outer
region of the torus is primarily ejected. As already pointed out in
Sec.~\ref{sec2.1} (see the upper panels of Fig.~\ref{fig2}), the
electron fraction for the outer part of the torus surrounding the
remnant massive neutron star is fairly high as $Y_e \agt 0.25$.  Thus,
the electron fraction for this ejecta component is typically 0.2--0.5
irrespective of $\alpha_{\rm vis}$~\cite{Fujiba17} (see the upper
right panel of Fig.~\ref{fig6}). That is, mildly neutron-rich matter
is ejected in contrast to the case of dynamical mass ejection.  In
particular, for the polar components with $\theta \alt 45^\circ$,
$Y_e$ is always larger than 0.3: see the upper panels of
Fig.~\ref{fig6}. Such ejecta can escape from the nucleosynthesis of an
appreciable amount of lanthanide elements~\cite{Fujiba17}, i.e., the
opacity is not enhanced.  Although the efficient heating source may
not be produced from the components of $Y_e \agt 0.35$~(see Fig.~5 of
Ref.~\cite{Wanajo14}), the fraction of such a component is minor for
this mass ejection mechanism. 
The typical ejecta velocity for this
component is $\sim 0.15$--$0.20c$ depending weakly on the value of
$\alpha_{\rm vis}$.


For the longer-term mass ejection with $t \agt 100$\,ms (up to $\sim
10$\,s), the viscous effects on the torus surrounding the central
massive neutron star play an important
role~\cite{FM13,FM14,FM15,Just2015,SM17}.  Broadly speaking, there are
two mechanisms for the mass ejection: one is viscosity-driven mass
ejection with neutrino irradiation and the other is late-time
viscosity-driven mass ejection.  Up to $\sim 1$\,s (i.e., for $100\,{\rm
ms} \alt t \alt 1\,{\rm s}$), matter ejected from the inner region of
the torus accounts for an appreciable fraction.  Because of the strong
neutrino heating effects near the massive neutron star, in particular in
the vicinity of its polar region, the mass ejection in the vicinity of
the rotational axis ($\theta \alt 30^\circ$) is activated: see middle
panels of Fig.~\ref{fig6}.  We refer to this mass ejection as
``viscosity-driven mass ejection with neutrino irradiation''.  For this
component, the mass ejection rate is $\sim 10^{-3}M_\odot/{\rm s}$ and
the typical velocity is $\sim 0.15c$ depending weakly on the values of
$\alpha_{\rm vis}$.  The neutron richness of this ejecta component is
not very high with $Y_e \agt 0.35$ irrespective of $\alpha_{\rm vis}$
because of the neutrino irradiation from the massive neutron star (e.g.,
Refs.~\cite{Perego,sekig15}). This indicates that this ejecta component
would be free from lanthanide elements~\cite{Oleg,Tanaka17,Fujiba17},
and hence, the opacity for this component would be small $\kappa \sim
0.1\,{\rm cm^2/g}$.  However, this ejecta component would not be a
strong heating source because heavy $r$-process elements (that are the
major heating sources) are not synthesized from the ejecta of $Y_e \agt
0.35$~\cite{Wanajo14,Fujiba17}\footnote{Ref.~\cite{Wanajo14} shows that 
the heating rate of dynamical ejecta with $Y_e\agt 0.35$  is suppressed by
a factor of $2$--$3$ at a few days. Note that, however, the heating rate of the
first r-process peak depends sensitively on the abundance pattern. Different 
values of the expansion time scale and entropy would result in higher heating
rates~\cite{Lippuner15}.}.  This type of the mass ejection process
continues as long as the massive neutron star (and torus) is present.

Our simulations for $\alpha_{\rm vis}=0.02$ and 0.04 were performed
for a long time scale 2--3\,s~\cite{Fujiba17} and it shows that the
longer-term mass ejection from the viscosity-driven expanding torus
occurs for $t \agt 1$\,s after the merger, but in a manner different
from the viscosity-driven mass ejection with neutrino irradiation: In
this late-time mass ejection mechanism, the matter is ejected
primarily toward the equatorial-plane direction ($\theta \agt
30^\circ$: see the bottom panels of Fig.~\ref{fig6}).  We refer to
this long-term mass ejection as ``late-time viscosity-driven mass
ejection from torus''.  The mass ejection rate for this ejection is
typically $\sim 10^{-2}M_\odot/{\rm s}$ depending weakly on the value
of $\alpha_{\rm vis}$, and it is enhanced earlier for the larger
values of $\alpha_{\rm vis}$: This mass ejection is initially
suppressed by the presence of the fall-back material that comes from
the failed-dynamical ejecta component, but after the density of the
fall-back material decreases, the mass ejection sets
in~\cite{Fujiba17}. For the larger values of $\alpha_{\rm vis}$, the
early viscosity-driven ejection helps to blow off a large fraction of
the fall-back material, and hence, this late-time viscosity-driven
mass ejection sets in earlier and in a higher ejection rate. Since the
mass ejection continues for seconds, the total ejecta mass in this
mechanism can be appreciably larger than $10^{-2}M_\odot$ (i.e.,
comparable to the torus mass) for large values of $\alpha_{\rm vis}
\agt 0.02$.

This late-time viscosity-driven mass ejection from the torus has been
already discovered by previous works~\cite{FM13,FM14,Just2015}, and
the mechanism is summarized as follows: For the late phase of the
evolution of the torus $\agt 1$\,s, neutrino cooling becomes
inefficient in the outer part of the torus with $r \agt 1000$\,km
because its temperature decreases to be low $\alt 1$\,MeV. Then, the
outer part of the torus expands by the viscous heating and viscous
angular momentum transport without appreciable cooling by neutrinos,
primarily toward the direction of the equatorial plane ($\theta \agt
45^\circ$). A part of the torus component of mass of order $10^{-2}
M_\odot$ is subsequently ejected from the system spending $\sim
10$\,s. For this component, the typical velocity is low, $\sim 0.05c$,
because the mass ejection occurs far from the central object (i.e.,
the typical velocity scale should be low). 

References~\cite{FM13,Just2015,SM17} focus on the case in which the
central object is a black hole and show that the value of $Y_e$ for
this ejecta component is unlikely to be very high for this case (see
the next subsection).  However, in the presence of a massive neutron
star that is the strong neutrino emitter, the value of $Y_e$ is
relatively high with 0.3--0.4 (compare histograms in the middle and
bottom panels of Fig.~\ref{fig6}), implying that the value of $\kappa$
is $\ll 10\,{\rm cm^2/g}$. Although the low velocity may prevent the
ejecta from shining in the early time of $t \alt$ a few days, the low
value of $\kappa$ may compensate this property (see
Sec.~\ref{sec3.1}).  Since the electron fraction is not very high for
this ejecta component in the presence of a massive neutron star, only
relatively light $r$-process elements are likely to be synthesized.
This suggests that the heating rate by the radioactive decay is
slightly lower than that by the heavier $r$-process elements~(see
Fig.~5 of Ref.~\cite{Wanajo14}).

For the early and long-term viscosity-driven ejecta, the typical
velocity is smaller than $0.2c$~\cite{Fujiba17}. Thus, the velocity is
slightly lower than that for the dynamical ejecta. This implies that
the dynamical ejecta would not be caught up by most part of the
viscosity-driven ejecta. A large fraction of the viscosity-driven
ejecta would be hidden by the dynamical ejecta, if we observe the
merger event from the direction of the binary orbital plane. However,
for the GW170817 event, the observer is likely to be located in a
polar region~\cite{LIGO817} and all the ejecta components could be
observed (see Sec.~\ref{sec3}).

To summarize, we find that the total ejecta mass could be $\agt
0.03M_\odot$ for a reasonable value of $\alpha_{\rm vis} \agt 0.02$
and the electron fraction for the ejecta is mildly neutron-rich in the
presence of a massive neutron star: $Y_e$ is distributed between 0.2
and 0.5 and a major fraction of the ejecta has a value of $Y_e$ larger
than 0.25.  In particular, for the long-term viscosity-driven ejecta
component, $Y_e$ is likely to be always larger than $\sim 0.3$. Thus,
in these ejecta components, the amount of lanthanide elements should
be quite small because for their nucleosynthesis, a sufficiently low
value of $Y_e \alt 0.25$ is required~\cite{Oleg,Tanaka17}. (See also
Table~\ref{tab3} for a summary for mass ejection mechanisms.)

In our study, we employ a model resulting from the merger of an
equal-mass binary neutron star as an initial condition.  In this
model, the torus mass is $\sim 0.2M_\odot$. In the presence of
asymmetry in mass of binaries, the torus mass would be slightly
larger, $\sim 0.3M_\odot$ (see Table~\ref{tab1}). For such cases, the
ejecta mass may be larger than the value estimated here. Studies for
such models are left for the future. We also note that for the DD2
EOS, the lifetime of the massive neutron star is quite long $\gg
1$\,s.  For the EOS in which the value of $M_{\rm max}$ is not as high
as that for the DD2 EOS, the massive neutron star could collapse to a
black hole within $\alt 1$\,s. Even for such EOSs, the viscosity-driven
mass ejection from the torus should continue after the black-hole
formation, but because of shorter neutrino-irradiation time, the value
of $Y_e$ is likely to be smaller than 0.3 in this scenario as
indicated in Ref.~\cite{FM14}. Studies for this case are also left for
the future.

\subsubsection{Evolution of black hole-torus system}\label{sec3.2.2}

To date, no detailed simulation for the evolution of black hole-torus
systems has been performed incorporating both general relativistic
gravity and neutrino heating together.  The simulations for this
system have been performed employing either viscous hydrodynamics in
a pseudo-Newtonian gravitational field with neutrino heating and
cooling~\cite{FM13,FM14,FM15,Just2015} or general relativistic
magnetohydrodynamics in a Kerr black-hole background with no
neutrino heating~\cite{SM17}.  We here summarize the results obtained
from these simulations. 

Simulations for spinning black hole-torus
systems~\cite{FM13,FM14,FM15,Just2015,SM17} with a high black-hole
spin of $\sim 0.8$ indicate that $\sim 20\%$ of the torus mass could
be ejected as a wind component through the long-term viscous process
in the torus. As already mentioned in the previous
subsection~\ref{sec3.2.1}, this viscosity-driven ejection takes place
for $t \agt 1$\,s after the temperature of the torus decreases to
$\alt 1$\,MeV.  All the previous simulations suggest that this
viscosity-driven ejecta is fairly neutron-rich with $Y_e=0.1$--0.5 and
with the peak at $Y_e=0.2$--0.3~\cite{FM13,FM14,FM15,Just2015,SM17}.
This indicates that the opacity for this component is likely to be as
high as that for the dynamical ejecta component because of the
nucleosynthesis of lanthanide elements.  Also shown is that the
velocity of this ejecta component is relatively low as 0.01--$0.1c$,
with the typical velocity $\sim 0.05c$, because the mass ejection from
the torus occurs in a region distant from the central region.  Thus,
this component may not be well suited for describing the early shining
of GW170817 (see Eq.~(\ref{tpeak})).

The mass ejection in this viscous process primarily proceeds to the
direction of the equatorial plane: Only a minor fraction of the mass is
ejected toward the polar direction.  The recent MHD simulation in
general relativity~\cite{SM17} indicates that the ejecta properties are
slightly modified by the MHD effect.  One effect observed in the MHD
simulation is to increase the fraction of the polar ejecta
component. However, the equatorial ejecta component is still dominant
over the polar one even in the MHD simulation, and overall, the ejecta
is always neutron-rich.

One of the concerns for these simulations is that initial conditions
might not be very realistic. The initial rotational profile was
typically given by assuming a constant specific angular momentum,
which is far from Keplerian and hence not very physical.  In
addition, $Y_e=0.1$ profile is initially given taking into account
that the accretion torus is neutron-rich because of its high density
resulting in a high degeneracy of electrons. However, in reality, the
outer part of the torus is likely to have a larger value of $Y_e$ up
to $\sim 0.4$ in the context of binary neutron star mergers (see the
upper panel of Fig.~\ref{fig1}). For this problem, more detailed
realistic simulations are awaited. However, in the following section,
we discuss possible scenarios based only on our results and those
reported in Refs.~\cite{FM13,FM14,FM15,Just2015,SM17}. 

\section{Models for GW170817}\label{sec3}

\subsection{Models for Macronova/kilonova}\label{sec3.1}

First, we summarize several approximate relations satisfied for the
macronova/kilonova model~\cite{Li,Metzger2010}.  The energy source in
this model is the radioactive decay of $r$-process elements. As the
radioactive heating rate declines monotonically with time, the
observed luminosity reaches the peak, $L_{\rm peak}$, on the photon
diffusion time scale of the ejecta (e.g., Ref.~\cite{Metzger2010}):
\beqn
t_{\rm peak} &\approx &
\sqrt{{ \xi \kappa M_{\rm ej} \over 4\pi c \bar v_{\rm ej}}} \nonumber \\
&\approx& 1.9 \,{\rm d} \, \xi^{1/2}
\left({\kappa \over 1\,{\rm cm^2/g}}\right)^{1/2}
\left({M_{\rm ej} \over 0.03M_\odot}\right)^{1/2} \nonumber \\
&& \hspace{3cm} \times\left({\bar v_{\rm ej} \over 0.2c}\right)^{-1/2}, 
\label{tpeak}
\eeqn
where $\xi$ is a parameter associated with the degree of asphericity
of the ejecta with $\xi \leq 1$, which depends on the geometry of the
ejecta.  
We note that the asphericity of the ejecta
profile can decrease $\xi$ along our line of sight.  However, the
degree of the asphericity for the dynamical and viscosity-driven
ejecta is $1/2 \alt \xi \leq 1$ (unless the binary mass asymmetry is
extremely high), and hence, its effect is not very
significant~\cite{Kasen17}.  Thus, the peak time is unlikely to be
significantly different from Eq.~(\ref{tpeak}). We note that it is
possible to consider that the velocity is enhanced effectively by
$\xi$ as $\bar v_{\rm ej}/\xi$. Thus, in the presence of the
asphericity, the effectively velocity can be increased if we consider
a model in the assumption of spherical symmetry.  We also note that
the peak luminosity and temperature can be enhanced by the asphericity
effect~\cite{Kyutoku13,Tanaka14a}; these effects should be taken into
account for detailed modeling.

Although the energy generation rate of beta decay in the
macronova/kilonova emission is robustly described as $\propto
t^{-1.3}$ for $t\gtrsim 1$\,d~\cite{Metzger2010,Hotoke17}, the thermalization
efficiency of decay products in the ejecta has to be taken into
account for the actual heating rate.  The specific heating rate for
the hypothetical abundance, in which the solar-abundance pattern is
assumed to be achieved, is given approximately as (e.g., see
Refs.~\cite{Tanaka14a,Hotoke16b})
\beq
\dot \varep \approx 1.6 \times 10^{10}\,{\rm erg/s/g}
\left({t \over {\rm day}}\right)^{-1.3}, \label{heat}
\eeq
when both electrons and gamma-rays are fully thermalized.  Gamma-rays
start leaking from the ejecta at $t_{{\rm in},\gamma}\approx 0.6\,{\rm
  d}\,(M_{\rm ej}/0.03M_{\odot})^{1/2} (\bar v_{\rm ej}/0.2c)^{-1/2}$,
where we used the inelasticity of the Compton scattering
\citep{Hotoke16b,Barnes16}.  The thermalization of electrons starts
being inefficient at $t_{{\rm in},e}\approx 18\,{\rm d} \, (M_{\rm
  ej}/0.03M_{\odot})^{1/2} (\bar v_{\rm ej}/0.2c)^{-3/2}$~\cite{Barnes16}.
For $t_{{\rm in},\gamma}\alt t \alt t_{{\rm in},e}$, the radioactive
heating is dominated by electrons, and the specific heating rate is
described approximately by
\beq
\dot \varep \approx 0.5 \times 10^{10}\,{\rm erg/s/g}
\left({t \over {\rm day}}\right)^{-1.3}.
\eeq
In addition to beta decay, alpha decay and spontaneous fission may
significantly enhance the heating rate at late times depending on the
abundance of heavy nuclei $A\geq 210$.  It is worthy to note that the
heating rate of alpha decay and fission arises, at late times, as the
shallower decline rate $\propto t^{-1}$ than that of beta
decay~\cite{Hotoke16b,Barnes16}. We also note that in the absence of
heavy $r$-process elements (like second- and third-peak elements), the
heating rate would be much lower than that shown here (see Fig.~5 of
Ref.~\cite{Wanajo14}).

For $t \agt t_{\rm peak}$, the total luminosity (in the hypothetical
presence of heavy r-elements) is given approximately by
\beqn
L &\approx& \dot \varep M_{\rm ej} \nonumber \\
&=&(0.3{\rm -}1.0) \times 10^{42}\,{\rm erg/s} 
\left({M_{\rm ej} \over 0.03M_\odot}\right)
\left({t \over {\rm day}}\right)^{-1.3}. \nonumber \\
\label{lumi}
\eeqn
We note that in the presence of other strong energy sources, e.g., a
magnetar central engine, the total luminosity may be higher than that
in Eq.~(\ref{lumi}), but in this section, we do not consider this
possibility.

\begin{figure}[t]
\begin{center}
\includegraphics[width=84mm]{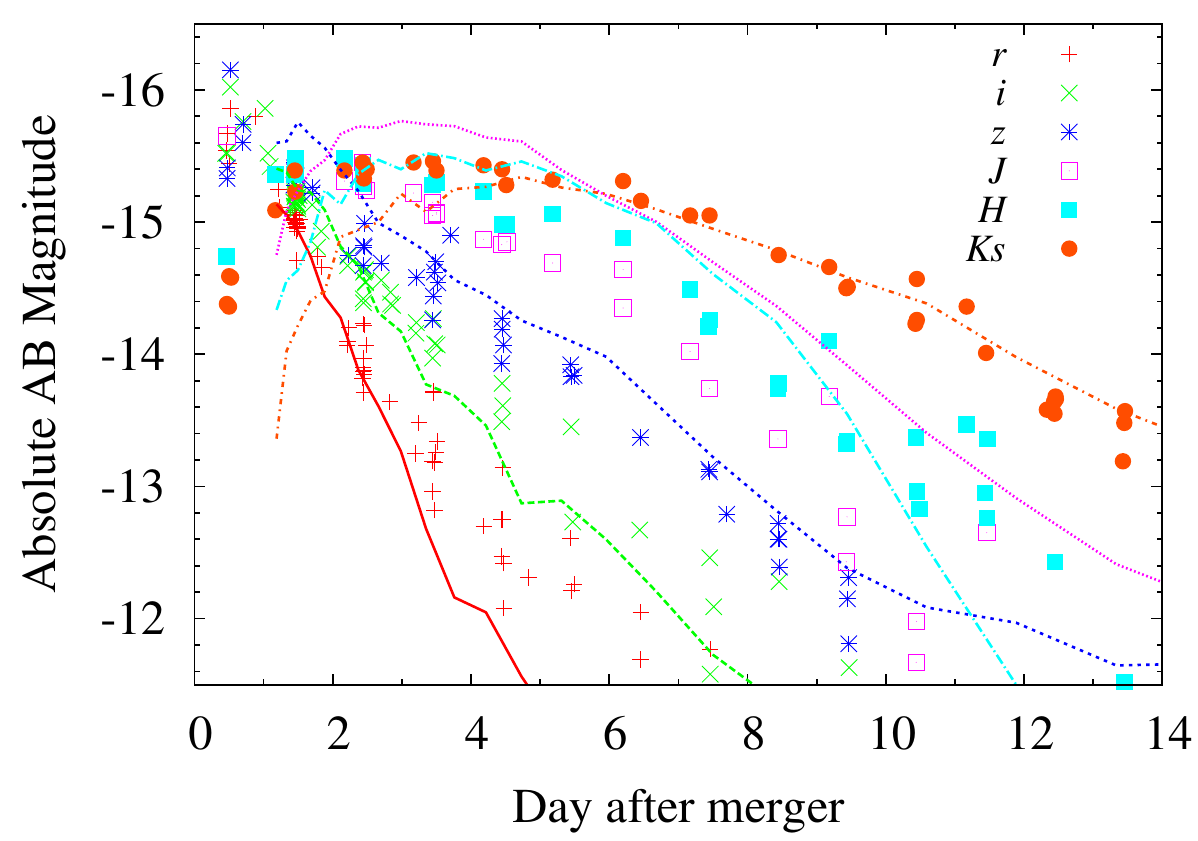}
\caption{Observational light curves (in terms of the data taken from
  Ref.~\cite{Villar}) and a light curve model~\cite{EM0} of the
  electromagnetic counterparts of GW170817. Plotted are the absolute
  AB magnitudes for the $r$, $i$, $z$, $J$, $H$, and $K_s$ bands. The
  horizontal axis shows the day spent after merger of the binary
  neutron stars. Here, we assume that the distance to the source is 40\,Mpc. 
\label{fig3}}
\end{center}
\end{figure}

Figure~\ref{fig3} shows observational light curves where the data are
taken from Ref.~\cite{Villar}. For describing the electromagnetic
counterparts of GW170817 for $t \alt 5$\,d, the following
observational results give the fundamental constraints to the free
parameters such as the ejecta mass, velocity, and opacity: (i) The
peak absolute (AB) magnitude in the $r$, $i$, and $z$ bands is
$\approx -16$~mag assuming that the distance to the source is 40\,Mpc
(the required luminosity for these bands is, broadly speaking, (3--5)
$\times 10^{41}\,{\rm erg/s}$) and the peak luminosity is reached
within $\sim 1$\,d after the merger.  (ii) The peak absolute (AB)
magnitude in the IR bands ($J$, $H$, and $K$ bands) is $\approx
-15.5$~mag (the required magnitude for these bands is approximately
$10^{41}\,{\rm erg/s}$ for the $J$ band and $3 \times 10^{40}\,{\rm
  erg/s}$ for the $K$ band), and this peak luminosity is reached in a
week after the merger.  Note that the observed spectrum is consistent
broadly with the black-body one with decreasing temperature (but see
Ref.~\cite{Mccully17} for detailed comparisons), and hence, the
evolution of the luminosity is consistent with the macronova/kilonova
model.

The early peak time for these observational results suggests that the
opacity cannot be as large as $\kappa=10\,{\rm cm^2/g}$ even for $\bar
v_{\rm ej} \sim 0.2c$ (see Eq.~(\ref{tpeak})). The high peak
luminosity also suggests that the ejecta mass should be appreciably
larger than $0.01M_\odot$ (see Eq.~(\ref{lumi})). The constraint,
$\kappa \ll 10\,{\rm cm^2/g}$, implies that the electromagnetic
counterpart should not contain a large amount of lanthanide elements
at least along our line of sight in the early time (for a few days
after the onset of merger). This strongly suggests that the ejecta
would be composed not only of dynamical ejecta but also of other
components like viscosity-driven-ejection components because the
dynamical ejecta primarily synthesizes heavy $r$-process elements
including lanthanide elements.  Also the high luminosity (i.e., high
ejecta mass $> 0.01M_\odot$) suggests that the ejecta would not be
composed only of dynamical ejecta.

In the late-phase of the electromagnetic counterparts of GW170817 with
$t \agt 5$\,d, a significantly reddening feature is found~(e.g.,
Refs.~\cite{EM3,EM6,Tan17}). For describing this component, the
opacity should be high $\kappa \sim 10\,{\rm cm^2/g}$, and hence, an
appreciable amount of the lanthanide synthesis is required. This
component is likely to be supplied from the dynamical ejecta and
viscosity-driven components obscured by the dynamical ejecta.

The solid curves of Fig.~\ref{fig3} denote a light curve
model~\cite{EM0} for the electromagnetic counterparts of
GW170817. This model assumes that the spherical ejecta expands in a
homologous manner with the average velocity $0.1c$ and with the mass
$M_{\rm ej}=0.03M_\odot$.  In this example, the opacity is determined
for a hypothetical abundance of $r$-process elements synthesized from
the ejecta of $Y_e=0.25$~\cite{Tanaka17} and results approximately in
$\kappa \sim 1\,{\rm cm^2/g}$.  This model approximately captures the
features for the observed event. We note that for a model in which
$M_{\rm ej}=0.03M_\odot$ and $\kappa=10\,{\rm cm^2/g}$, the $i$-band
luminosity at $t=1$\,d is only $\approx -15$\,mag, and moreover, the
peak time for $H$-band is delayed significantly to $t_{\rm peak} \agt
5$\,d~\cite{Tanaka17}.  These results suggest that the low value of
$\kappa$ is one of the keys for interpreting the observational results
of GW170817.

Paying particular attention to the two constraints (i) and (ii), we here
explore the following two scenarios for interpreting the GW170817 event:
One scenario is based on the numerical results with the SFHo EOS, and
the other is based on the results with the DD2 EOS.  For the given
constraint to the total mass of the binary neutron stars of GW170817, $m
\geq 2.73M_\odot$, in the former, the remnant is a spinning black hole
surrounded by a torus, and in the latter, it is a long-lived massive
neutron star surrounded by a torus.  In the following subsections, we
finally conclude that (I) the current numerical-relativity simulations
do not support the SFHo model in which a black hole is formed in a short
time scale after the onset of merger and hence long-term strong sources
for the neutrino irradiation may be absent in the merger remnant
(because of the same reason~\cite{Francois,kyutoku}, the black
hole-neutron star model for GW170817 is likely to be rejected by the
observation of the electromagnetic counterparts): (II) the presence of a
long-lived remnant massive neutron star found in the DD2 model, which is
a long-term strong emitter of neutrinos and is suitable for increasing
the electron fraction of the ejecta, is favorable for interpreting the
observational results of GW170817.

\begin{table*}[t]
\centering
\caption{\label{tab3} Summary of mass ejection mechanisms. Table shows
  ejection type, ejecta mass, typical velocity, electron fraction,
  major direction of the mass ejection, and ejection duration are
  summarized for the SFHo and DD2 models. $t_\nu$ and $t$ denote the
  duration of the neutrino emission and the time after the onset of
  merger, respectively.  We note that for the EOS in which the value
  of $M_{\rm max}$ is not as high as that for the DD2 EOS, the massive
  neutron star could collapse to a black hole within $\alt 1$\,s. Even
  for such EOSs, the viscosity-driven mass ejection from torus should
  continue after the black-hole formation, but because of shorter
  neutrino-irradiation time, the value of $Y_e$ is likely to be
  smaller than 0.3.  }
\begin{tabular}{cccccc}
\hline\hline
&SFHo model  & & & &
\\ \hline
Type of ejecta & Mass $(M_\odot)$ & $\bar v_{\rm ej}/c$ & $Y_e$ & ~Direction~ 
& Duration 
\\ \hline
Dynamical ejecta & $\sim 10^{-2}$ & $\sim 0.2$ & 0.05--0.5 
& $\theta \agt 45^\circ$ 
& $t \alt 10$\,ms
\\ 
Viscosity-driven ejecta from torus& (1--2)$\times 10^{-2}$ 
& 0.01--0.1 & 0.1--0.5 & $\theta \agt 45^\circ$ & $t \sim 1$--10\,s
\\ \hline \hline 
& DD2 model & & & & 
\\ \hline
Type of ejecta & Mass $(M_\odot)$ & $\bar v_{\rm ej}/c$ & $Y_e$ & Direction 
& Duration 
\\ \hline
Dynamical ejecta & $O(10^{-3})$ & $\sim 0.2$ & 0.05--0.5 & $\theta \agt 45^\circ$ 
& $t \alt 10$\,ms
\\ 
Early viscosity-driven ejecta & ~$\sim 10^{-2} (\alpha_{\rm vis}/0.02)$~
& ~0.15--0.2~ & 0.2--0.5 & ~$\theta \agt 30^\circ$~ & ~$t \alt 100$\,ms~
\\ 
~~~Viscosity-driven ejecta with neutrino irradiation~~~ & ~$t_\nu
\times10^{-3}$/s~
& $\sim 0.15$ & 0.35--0.5 & $\theta \alt 30^\circ$ & ~$t \alt t_\nu \sim 10$\,s~
\\ 
Late-time viscosity-driven ejecta from torus& $>10^{-2}$ 
& $\sim 0.05$ & 0.3--0.4 & $\theta \agt 30^\circ$ & $t \sim 1$--10\,s
\\
\hline\hline
\end{tabular}
\end{table*}

For the help of understanding each model, in Table~\ref{tab3}, we
summarize the type of the ejecta and properties of each ejecta
component for the SFHo and DD2 models separately.

\subsection{Scenario for the soft EOS}\label{sec3.2}

\begin{figure}[t]
\begin{center}
\includegraphics[width=80mm]{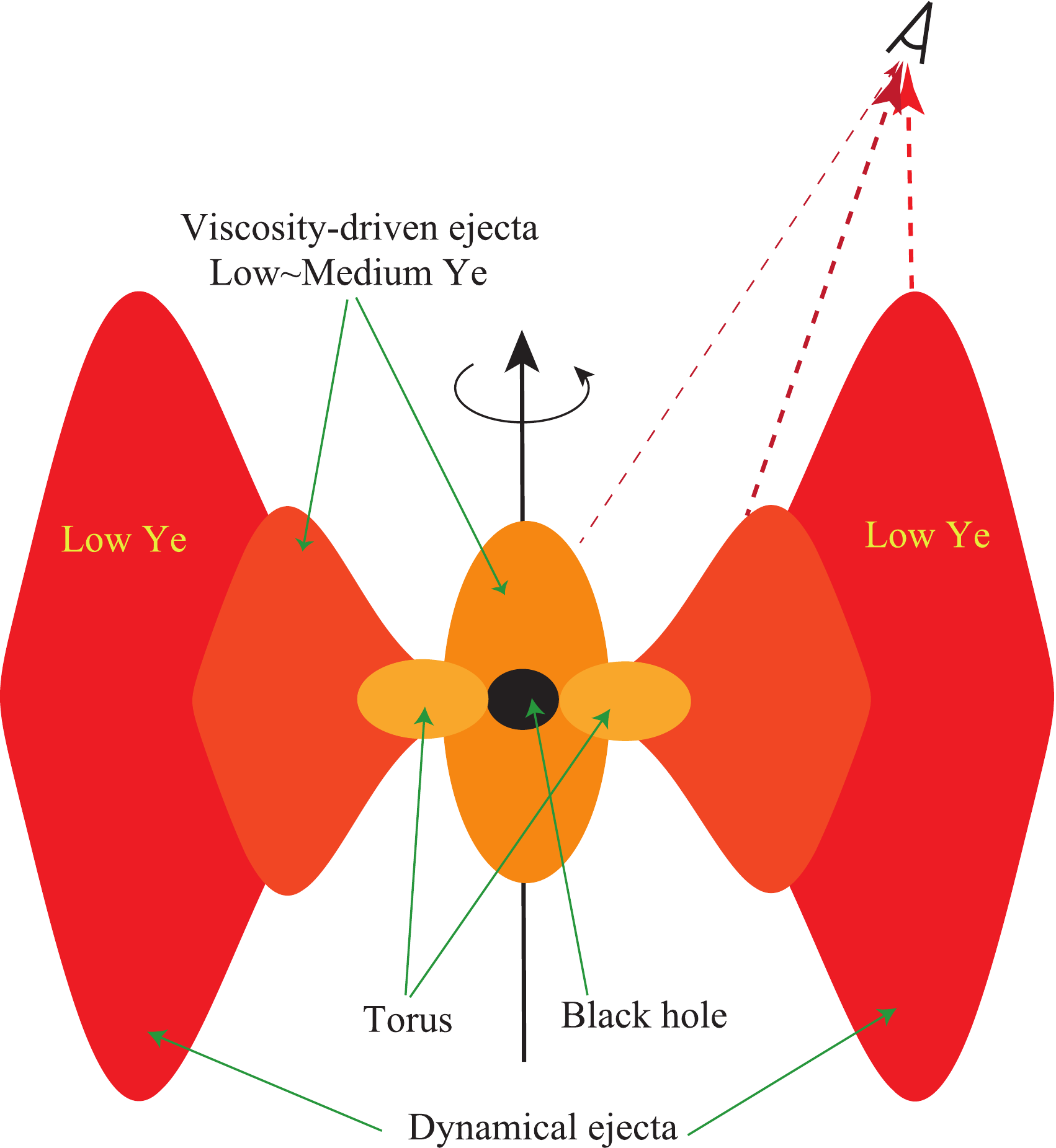}
\caption{Schematic picture of the ejecta profile for the case of a
  soft EOS in which a black hole is formed in $\sim 10$\,ms after the
  onset of merger. The largest anisotropic-shell component (red color)
  denotes the dynamical ejecta.  The smaller anisotropic-shell (red)
  and polar components (ocher) denote the viscous/MHD ejecta from the
  torus, respectively. The ``Low $Y_e$'' implies that it contains
  neutron-rich matter with $Y_e \alt 0.2$, which synthesizes an
  appreciable amount of lanthanide elements and contributes to
  enhancing the opacity to $\kappa \sim 10\,{\rm cm^2/g}$. The polar
  component could have ejecta of $Y_e=0.3$--0.4 but it is a minor
  component in mass.  The black filled circle and neighbouring
  (orange) ellipsoids in the central region denote a spinning black
  hole and accretion torus surrounding the black hole,
  respectively. Since the opacity is entirely high for all the major
  ejecta components, it is difficult to describe the observational
  results (in particular early peak time) for the electromagnetic
  counterparts of GW170817 by this model.
\label{fig4}}
\end{center}
\end{figure}

First, we describe the scenario based on the numerical-relativity
results for the SFHo EOS as a model that is not well suited for
interpreting the observations for the electromagnetic counterparts of
GW170817.  As shown in Sec.~\ref{sec2.1}, in this EOS model, the
merger with $m=2.7$--$2.8M_\odot$ results in temporal formation of a
hypermassive neutron star and it collapses, in $\alt 10$\,ms after its
formation, to a spinning black hole of dimensionless spin $\sim 0.7$
surrounded by a torus of mass $\sim 0.1M_\odot$. Because the lifetime
of the hypermassive neutron star is shorter for the higher total mass
of the system, we here assume that the lifetime would be $\alt 5$\,ms
taking into account the total mass of GW170817, $m \agt 2.73M_\odot$
(i.e., we assume that the lifetime would be shorter than the viscous
time scale in the hypermassive neutron star, written in
Eq.~(\ref{eq:mnsvis})). In this model, the mass of the dynamical
ejecta is $\sim 0.01M_\odot$ and this dynamical ejecta always contains
a substantial fraction of neutron-rich elements with $Y_e\alt 0.2$
irrespective of the total mass and mass ratio.  Thus, the opacity of
the dynamical ejecta is high as $\kappa \sim 10\,{\rm
  cm^2/g}$~\cite{BK2013,TH2013,opacity,Tanaka17} due to the existence
of an appreciable amount of lanthanide elements~\cite{Oleg,Tanaka17}.
The dynamical ejecta has a quasi-isotropic shell structure (see
Fig.~\ref{fig4}).

For this model, the remnant black hole is surrounded by a torus of
mass 0.05--$0.1M_\odot$.  Simulations for spinning black hole-torus
systems~\cite{FM13,FM14,FM15,Just2015,SM17} indicate that $\sim 20\%$
of the torus mass could be ejected as a viscosity-driven component
through the long-term viscous process in the torus. This suggests that
by the viscous process, matter with mass of $\sim 0.01$--$0.02M_\odot$
could be ejected. However, these previous simulations suggest that
this viscous component is fairly neutron-rich of a wide distribution
of $Y_e=0.1$--0.5 with the peak at $Y_e=0.2$--0.3. From such ejecta,
an appreciable amount of lanthanide elements should be
synthesized~\cite{Oleg,Tanaka17}.  This indicates that the opacity for
this component is likely to be as high as that for the dynamical
ejecta component.  Also shown is that the mass ejection in this
viscous process primarily proceeds to the direction of the equatorial
plane (not to the polar direction).  The typical velocity of this
ejecta component is lower than that for the dynamical ejecta and in
addition its morphology is similar to that of the dynamical ejecta
(for which the mass ejection occurs also primarily in the direction of
the binary orbital plane).  All these facts suggest that the time
scale to reach the peak luminosity is likely to be much longer than
1\,day due to the high opacity and $\bar v_{\rm ej} \leq 0.25c$ even
for $\xi \sim 1/2$ (see Eq.~(\ref{tpeak})): This model is not suitable
for reproducing the optical--IR counterparts for GW170817. We speculate
that this conclusion may be universal for any EOS model in which a
long-lived massive neutron star is not formed as the merger remnant.
For examining this speculation, we need to perform more simulations
employing different EOSs.

If significant viscosity-driven mass ejection could occur in the polar
direction with high velocity and with moderate neutron richness as
$Y_e \agt 0.25$, this model could be viable. Such ejection may be
possible if significant neutrino heating occurs from the inner
edge of the torus around a spinning black hole.  Indeed, general
relativistic simulations of Ref.~\cite{SST07} (with no neutrino
heating) suggest that strong neutrino emission with a luminosity of
appreciably higher than $10^{52}\,{\rm erg/s}$ may be possible from a
torus of mass $\agt 0.1M_\odot$ surrounding a spinning black hole of
dimensionless spin 0.75. This possibility deserves more detailed
exploration.  For this purpose, we need a detailed numerical work
incorporating neutrino heating and general relativity for the
evolution of the merger remnant.

\subsection{Scenario for the stiff EOS}\label{sec3.3}

\begin{figure}[t]
\begin{center}
\includegraphics[width=80mm]{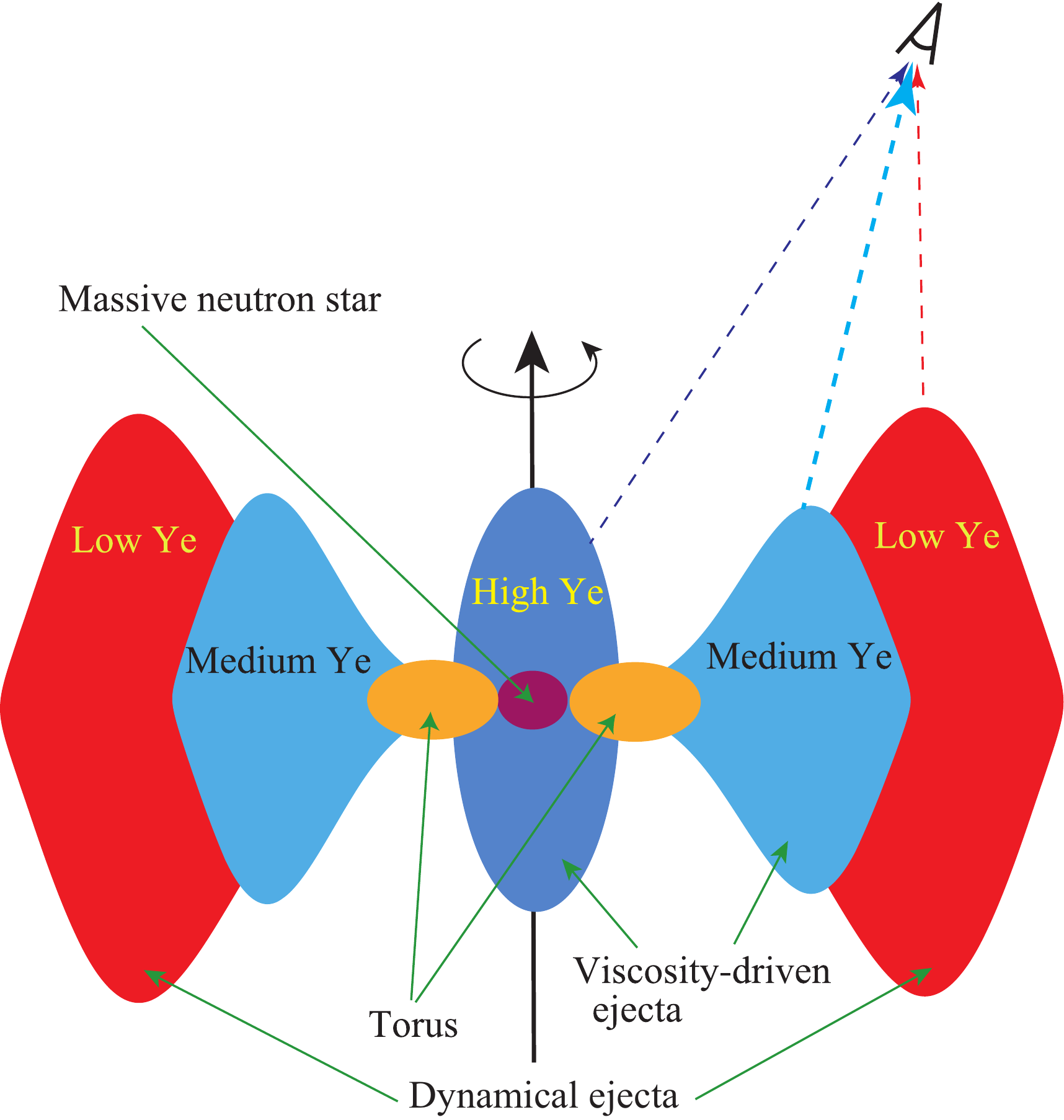}
\caption{Schematic picture of the ejecta profile for the case of a
  stiff EOS in which a long-lived massive neutron star is formed as a
  remnant. The largest anisotropic-shell component (red color) denotes
  the neutron-rich dynamical ejecta.  The smaller anisotropic-shell
  component (blue color) denotes the early viscosity-driven ejecta and
  long-term viscosity-driven ejecta from the torus.  The polar
  spheroid component (dark blue color) denotes the viscosity-driven
  ejecta from the torus influenced by neutrino irradiation from the
  massive neutron star.  The ``Low $Y_e$'' implies that it contains
  neutron-rich matter with $Y_e \alt 0.2$, which contributes to
  enhancing the opacity through the nucleosynthesis of lanthanide
  elements.  The ``Medium $Y_e$'' and ``High $Y_e$'' imply that it
  does not contain such neutron-rich matter because $Y_e \agt 0.25$
  and $Y_e \agt 0.35$, respectively.  The filled (purple) circle and
  neighbouring small (orange) ellipsoids in the central region denote
  a massive neutron star and accretion torus surrounding it. We note
  that the ``Low $Y_e$'' component has high average expansion velocity
  of $\bar v_{\rm ej}\sim 0.2c$ while the ``Medium'' and ``High''
  components have slower velocity, 0.1--$0.2c$. Note that the
  gravitational-wave observation indicates that we observe the merger
  remnant of GW170817 along the direction of $\theta \leq 28^\circ$
  from the rotation axis.
\label{fig5}}
\end{center}
\end{figure}

As shown in Sec.~\ref{sec2.1}, in the stiff EOS model like DD2, the merger
for $m=2.7$--$2.8M_\odot$ results in the formation of a long-lived
massive neutron star, which is differentially rotating at its 
formation. The remnant massive neutron star is surrounded by a dense
torus of mass $\sim 0.2$--$0.3M_\odot$.  

For this model, the mass of the dynamical ejecta is of order
$10^{-3}M_\odot$. This ejecta contains a sufficient fraction of
low-$Y_e$ component, and hence, a significant amount of lanthanide
elements are synthesized, resulting in a high opacity
$\kappa \sim 10\,{\rm cm^2/g}$.  The dynamical ejecta has an
anisotropic-shell structure in this EOS (see Fig.~\ref{fig5}). The
typical ejecta velocity is $\bar v_{\rm ej} \sim 0.2c$ or slightly slower. 

Because the remnant massive neutron star is initially differentially
rotating, the subsequent mass ejection is likely to be induced by the
viscous effects (i.e., early viscosity-driven mass ejection).  As
described in Sec.~\ref{sec2.2}, the degree of differential rotation of
the remnant neutron star decreases with time and it approaches a
rigidly rotating state on a time scale of $\sim 10$--20\,ms with a
reasonable value of $\alpha_{\rm vis}$ (see
Eq.~(\ref{eq:mnsvis})). During the transition of this rotating state,
matter is likely to be ejected. This mass ejection occurs in a fairly
anisotropic manner and the typical ejecta velocity is $\bar v_{\rm
  ej}=0.15$--$0.20c$, i.e., slightly smaller than that of the
dynamical ejecta.  The neutron richness of this ejecta component is
mildly high, i.e., $Y_e\approx 0.2$--0.5~\cite{Fujiba17}. What is nice
in this ejection is that for the high latitude ($\theta \alt
45^\circ$), the fraction of neutron-rich matter is small (see
Figs.~\ref{fig6} and~\ref{fig5}).  If the turbulent state of the
remnant massive star is sufficiently enhanced and the resulting
effective viscous parameter is sufficiently large as $\alpha_{\rm
  vis}\agt 0.02$, the ejecta mass in this mechanism could be $>
0.01M_\odot$.

Since the torus surrounding the central massive neutron star is also
differentially rotating, the viscosity-driven mass ejection from the
torus occurs for a long time scale of $\sim 1$--10\,s following the
early viscosity-driven ejection. For $100\,{\rm ms} \alt t \alt 1$\,s,
this mass ejection proceeds primarily toward the polar direction
because of the strong neutrino heating near the remnant massive
neutron star (viscosity-driven mass ejection with neutrino
irradiation)~\cite{Fujiba17}.  The typical ejecta velocity is $\bar
v_{\rm ej}=0.1$--$0.2c$ depending weakly on the value of $\alpha_{\rm
  vis}$.  The neutron richness of this ejecta component is not high,
$Y_e \agt 0.35$, because of the strong neutrino irradiation from the
remnant neutron star, and hence, the heating rate is sensitive to the
elemental abundance pattern, as already mentioned in
Sec.~\ref{sec2.2}.

In the later phase of the viscosity-driven mass ejection ($t > 1$\,s),
the mass is primarily ejected in a weakly anisotropic manner with the
average mass ejection rate of $10^{-2}M_\odot$/s and with low velocity
$\sim 0.05c$ (late-time viscosity-driven mass ejection from the
torus). For this component, the value of $Y_e$ is 0.3--0.4. (If the
lifetime of the massive neutron star is shorter than $\sim 1$\,s, the
value of $Y_e$ may be smaller than 0.3.)  Thus, we may expect a weak
lanthanide contamination and the presence of a relatively strong
heating source (not as strong as by the heavier $r$-process
elements~\cite{Wanajo14}).  This mass ejection from the expanding
torus is likely to continue for $\sim 10$\,s as found in
Ref.~\cite{FM14}, and the ejecta mass also could be of order
$0.01M_\odot$.

All these discussions (based mainly on our numerical-relativity
simulations) suggest that in this model, the mass of the mildly
neutron-rich viscosity-driven ejecta with the velocity 0.05--$0.15c$
could be $\agt 0.03M_\odot$ in total for $\alpha_{\rm vis} \agt 0.02$.
Since most of these viscosity-driven ejecta are not highly
neutron-rich with $Y_e \agt 0.25$, and thus, the nucleosynthesis of
lanthanide elements would be suppressed, their opacity is likely to be
$\kappa \sim 1\,{\rm cm^2/g}$~\cite{Oleg,Kasen15,Tanaka17}. In
particular for the ejected matter located for the high latitude
($\theta \alt 45^\circ$), $Y_e$ is always high (see
Figs.~\ref{fig6} and~\ref{fig5}).  This indicates that if an observer
is not located near the binary orbital plane, the effect of the
lanthanide curtain provided by the dynamical ejecta could be
avoided. Then, if the mass of the viscosity-driven ejecta is
sufficiently high as $\agt 0.03M_\odot$ (i.e., $\alpha_{\rm vis}$ is
sufficiently large $\sim 0.02$--0.04), the electromagnetic
observations for GW170817 can be naturally interpreted.

One unclear point in the early viscosity-driven ejection is that we do
not know whether $\alpha_{\rm vis}$ is really sufficiently large $\sim
0.02$--0.04 around the central region of the remnant massive
  neutron star, i.e., a sufficiently strong turbulence state is
realized or not there, although $\alpha_{\rm vis}=O(0.01)$ is a
reasonable magnitude for turbulent fluids: Indeed, our latest
  high-resolution MHD simulation~\cite{Kiuchi17} shows that at least
  for an outer region of the remnant massive neutron star and torus,
  $\alpha_{\rm vis}$ is likely to be enhanced to $\sim 0.02$.  To
assess the validity of this scenario, however, we need to perform a
high-resolution MHD simulation for the merger and post-merger of
binary neutron stars, in which several MHD instabilities such as
Kelvin-Helmholtz and magneto-rotational instabilities are well
resolved.  We note that if the initial torus mass of the merger
remnant is higher (e.g., for the merger of significant binary mass
asymmetry), the ejecta mass of $\sim 0.03M_\odot$ may be achieved for
a smaller value of $\alpha_{\rm vis}$. Thus, the required value for
$\alpha_{\rm vis}$ may be smaller.

In this section, we have paid particular attention to the optical-IR
counterparts in the relatively early phase of $\alt 5$\,days.  In the
late phase, the effect of the dynamical mass ejection of low $Y_e$
(i.e., of high values of $\kappa$) should be visible. The late-time
reddening~\cite{EM3,EM6,Tan17} is likely to be associated with the
dynamical ejecta component in our scenario.

\section{Discussion}\label{sec4}

\subsection{Perspective for constraining the neutron-star EOS through measuring tidal deformability}

In Sec.~\ref{sec3}, we proposed a model of the binary neutron star
merger suitable for interpreting the observational results for the
electromagnetic signals of GW170817. Our analysis suggests that the
neutron-star EOS would be stiff enough (i.e., the maximum mass for
cold spherical neutron stars is large enough) to produce a long-lived
massive neutron star after the merger for the total mass $m \agt
2.73M_\odot$. However, this suggestion primarily constrains the
maximum mass of cold neutron stars, not neutron-star radius.

One of the most promising methods to narrow down the possible EOS
candidates by constraining the typical radius of neutron stars is to
measure the tidal deformability of neutron stars through the
gravitational-wave observation of the late inspiral signals of binary
neutron stars~(e.g.,
Refs.~\cite{damour12,Wade14,Bernuzzi14,Aga15,Hotoke16}). For an event
of $S/N \approx 30$ to LIGO O2 sensitivity (for which the sensitivity
for a high-frequency band $\agt 400$\,Hz is not as good as for the
lower-band~\cite{170814}), the binary dimensionless tidal
deformability, $\Lambda$, would be distinguished up to $\delta \Lambda
\approx 400$ at 2-$\sigma$ level by analyzing gravitational waves from
binary neutron stars in close orbits~\cite{damour12,Hotoke16}.  Here,
$\Lambda$ is defined by
\beqn
\Lambda={8 \over 13} \biggl[
 && (1 + 7\eta -31 \eta^2)(\Lambda_1 + \Lambda_2)  \nonumber \\
&&  -\sqrt{1-4\eta}(1+9\eta-11\eta^2)(\Lambda_1-\Lambda_2) \biggr],  
\eeqn
and $\Lambda_1$ and $\Lambda_2$ are each dimensionless tidal
deformability in binaries. It is known that for a given value of the
chirp mass, $\Lambda$ depends very weakly on mass ratio (see, e.g., the
last five data in each raw of Table~\ref{tab2})~\cite{Jim17}.  The
gravitational-wave observation of GW170817 preliminary suggests that
$\Lambda$ is smaller than $\sim 800$ as the 90\% credible upper limit
for a hypothesis that the dimensionless spin parameter of neutron stars
is smaller than 0.05~\cite{LIGO817}. Thus, the DD2 EOS is marginally
acceptable, although the template employed in this preliminary analysis
tends to indicate a value of $\Lambda$~\cite{Hotoke16} larger than the
true one.  The error size for the observational result of GW170817 is
consistent with the analysis of Refs.~\cite{damour12,Hotoke16}. We note
that this observational result suggests that the neutron-star radius of
mass $1.35M_\odot$ should be smaller than $\sim 13$\,km at the 90\%
credible upper limit (see, e.g., Table~\ref{tab2}).

For the SFHo EOS in which the typical neutron-star radius is $R \sim
12$\,km, the maximum mass of cold spherical neutron stars is $M_{\rm
  max} \approx 2.06M_\odot$. For such a type of EOS, long-lived
massive neutron stars cannot be formed after the merger for $m \agt
2.7M_\odot$. However, if $M_{\rm max}$ is appreciably larger than
$2M_\odot$ for an EOS of $R \alt 12$\,km due to significant stiffening
of the EOS for the supra-nuclear-density region like in the EOS of
Ref.~\cite{Togashi} (for which $M_{\rm max} \approx 2.21M_\odot$ and
$R \sim 11.5$\,km), a long-lived massive neutron star may be formed
after the merger for $m \agt 2.73M_\odot$.  The maximum mass cannot be
increased arbitrarily because the sound speed has to be always smaller
than the speed of light.  However, detailed analyses of spherical
neutron stars~\cite{Jim} show that the maximum mass can be as high as
$\sim 2.2$--$2.3M_\odot$ even for the neutron-star radius of
11--12\,km.  The gravitational-wave and electromagnetic observations
for GW170817 suggest that such an EOS is a candidate even if 
the typical radius is small.

\subsection{Possible constraint to the neutron-star EOS through the observations of electromagnetic counterparts}

A possible constraint on the neutron-star EOS is obtained from the
absence of observational evidence for the existence of a rapidly
rotating magnetar remnant, which can release its rotational kinetic
energy:
\beqn
T_{\rm rot} & \approx & 1.1\times 10^{53}\,{\rm erg}
\left({M_{\rm MNS} \over 2.5M_\odot}\right)
\left({R \over 15\,{\rm km}}\right)^2 \nonumber \\
&& \hspace{2.5cm} \times
\left({\Omega \over 7000\,{\rm rad/s}}\right)^2, \label{trot}
\eeqn
where we used $T_{\rm rot}=I\Omega^2/2$ and $I=0.4M_{\rm
  MNS}R^2$~\cite{FIP86} with $I$ the moment of the inertia and $M_{\rm
  MNS}$ the mass of the massive neutron star.  As mentioned in
Sec.~\ref{sec3.2.1}, the remnant massive neutron star is likely to be
strongly magnetized due to several amplification processes of the
magnetic-field strength during the
merger~\cite{Kiuchi14,Kiuchi15,Kiuchi17}. If a force-free dipole
magnetic field with a strong magnetic-field like in
magnetars~\cite{magnetar} is established outside the merger remnant,
the system would release its rotational kinetic energy through strong
magnetic dipole radiation with luminosity~\cite{ST}:
\beqn
L_{\rm mag}&\approx& {B_p^2 R^6 \Omega^4 \over 6c^3} \nonumber \\
&\approx &1.7 \times 10^{50}\,{\rm erg/s}
\left({B_p \over 10^{15}\,{\rm G}}\right)^2
\left({R \over 15\,{\rm km}}\right)^6 \nonumber \\
&& \hspace{2.5cm} \times
\left({\Omega \over 7000\,{\rm rad/s}}\right)^4,
\eeqn
where $B_p$ is the magnetic-field strength at the polar region, $R$ is
the typical radius, and $\Omega$ is the angular velocity of the
remnant massive neutron star, respectively.  Thus, the spin-down time
scale of the remnant massive neutron star defined by $T_{\rm
  rot}/L_{\rm mag}$ is estimated as
\beqn \tau_B &&\approx 650\,{\rm s} \left({B_p \over 10^{15}\,{\rm
    G}}\right)^{-2} \left({M_{\rm MNS} \over
  2.5M_\odot}\right)\nonumber \\ && \hspace{1.5cm} \times \left({R
  \over 15\,{\rm km}}\right)^{-4} \left({\Omega \over 7000\,{\rm
    rad/s}}\right)^{-2}.  
\eeqn
As this spin-down process occurs on a time scale during which the
ejecta is still optically thick, the remnant magnetar produces a hot
bubble inside the ejecta, that accelerates the ejected matter. Thus, the
rotational kinetic energy is converted to the ejecta's kinetic energy.
If a substantial fraction of $T_{\rm rot}$ is injected into the
ejecta, we expect to observe (i) the expansion velocity of $\bar v_{\rm
  ej}\approx c$ (because $T_{\rm rot} \agt M_{\rm ej}c^2$), and  
(ii) a very bright radio, optical, X-ray signals
\citep{Metzger2014,MB2014, Horesh16}.  However, the electromagnetic
observations for GW170817 did not show any evidence of these
features. One possible interpretation for this is that such a strong
dipole magnetic field would not be established outside the remnant on
a time scale of a month, although the magnetic field inside the
massive neutron star is very strong $\agt 10^{16}$\,G. However, this
is not very natural, because there are many neutron stars of strong
magnetic fields in nature~\cite{magnetar}.

Another interpretation is that the massive neutron star collapses to a
black hole before a substantial fraction of the rotational kinetic
energy is released.  This condition is satisfied for the case that the
remnant object is a hypermassive neutron star. However, it is
difficult to interpret the fast rise of the optical--IR light curves
if the lifetime of the hypermassive neutron star is too short, e.g.,
the SFHo EOS case, as discussed in Sec.~\ref{sec3}.  A more favored
scenario is that the hypermassive neutron star collapses in a time
scale of the neutrino cooling of $\sim 10$\,s $\ll \tau_B$. If
high-luminosity gamma-rays detected by Fermi and INTEGRAL came from a
black hole surrounded by a torus, the collapse to the black hole may
occur at $\alt 1$\,s.  This scenario could be satisfied in the case of
a stiff EOS.  If this is the case, the maximum mass of the neutron
star could be constrained.  For instance, the gravitational mass of
the remnant massive neutron star at the collapse is likely to be $\sim
(2.60 \pm 0.05)M_\odot$ for the total mass of the binary for GW170817,
$m=2.73$--$2.78M_\odot$, because the gravitational-wave emission
(primarily during the inspiral phase), the long-term neutrino emission
(in the post-merger phase), and the mass ejection reduce the mass of
the system by $\sim (0.15 \pm 0.03)M_\odot$ in total (supposing that
the lifetime of the hypermassive massive neutron stars is $\sim
0.5$--2\,s). We suppose that at the onset of the collapse to a black
hole, the gravitational mass of the remnant massive neutron star in
rapid rotation would be such values.  Because the rapid and rigid
rotation increases the maximum-allowed mass for neutron stars by $\sim
0.4M_\odot$~\cite{CST94}, the expected maximum mass for cold spherical
neutron stars would be $\sim 2.15$--$2.25M_\odot$ (i.e., the maximum
mass for the DD2 EOS is slightly larger than the required value). This
is a reasonable value for typical stiff EOSs (see also Ref.~\cite{Marg}
for an independent analysis).

\subsection{Implications of GW170817 event for $r$-process nucleosynthesis}

Our model with the DD2 EOS shows that the total mass of ejected heavy
$r$-process elements with mass number $A \agt 90$ (i.e., the so-called
second and third peak elements) would be several$\times
10^{-3}M_\odot$--$10^{-2}M_\odot$. In our model, such $r$-process
elements are likely to be synthesized partly from the dynamical ejecta
and primarily from the early viscosity-driven
ejecta~\cite{Fujiba17}. For the latter, the $r$-process elements only up
to the second peak would be formed~\cite{Fujiba17}, but because the
total ejected mass by this mechanism dominates over that of the
dynamical ejecta, it will contribute to the majority of the total mass
of $r$-process elements. On the other hand, the dynamical ejecta will
contribute primarily to synthesis of the third-peak and lanthanide
elements because its neutron richness is quite
high~\cite{sekig15,sekig16}.

Assuming that the solar abundance~\cite{goriery} gives the mean values
for stars in the Galactic disk, the total mass of heavy $r$-process
elements with $A \geq 90$ in our Galaxy is approximately $5 \times
10^3M_\odot$. These elements also indicate a uniform abundance pattern
in metal-poor stars~\cite{Sneden}. This suggests that they are
synthesized in a single kind of the phenomenon. As discussed in
Ref.~\cite{Hotoke15}, mergers of neutron-star binaries (binary neutron
stars and black hole-neutron star binaries) are among the most
promising candidates for the source of the $r$-process
nucleosynthesis. Here, we assume that the binary neutron star mergers
are the dominant nucleosynthesis sources. Since GW170817
  indicates that the neutron-star radius is fairly
  small~\cite{LIGO817} and hence tidal disruption of neutron stars by
  black holes becomes less likely, this may be now a reasonable
  assumption.

If the $r$-process nucleosynthesis has occurred uniformly in the history
of our Galaxy, the merger rate of binary neutron stars is estimated to
be
\beq
10^{-4}\,{\rm yr}^{-1} \left({M_{A \geq 90} \over 5\times 10^{-3}M_\odot}
\right),
\eeq
where $M_{A \geq 90}$ denotes the average total mass of $r$-process
elements with $A \geq 90$ that are synthesized in one merger event. This
event rate may be slightly larger than the latest estimates such as that
based on a statistical study of observed binary neutron stars in our
Galaxy~\cite{Kim}. However, the detection of GW170817 by detectors of
the horizon distance of $D_{\rm eff} \approx 100$\,Mpc in half a year
observation with the duty cycle of $\sim 50$\% suggests that the event
rate of the binary neutron star mergers may be $\sim 10^{-4}$/yrs in a
Milky-way equivalent galaxy \cite{LIGO817}.  Therefore, the merger rate
estimated here is a reasonable value.

\subsection{Radio flares}

The (sub-relativistic) merger ejecta sweep up the interstellar matter
and form blast waves.  In the shocked matter, the magnetic fields are
amplified and electrons are accelerated.  This process will produce a
synchrotron radio flare~\cite{NP2011}.  The ejecta discussed in this
paper can be the source of the observable radio flares.

The radio flare will reach peak luminosity when the total swept-up
mass approaches the ejecta mass~\cite{NP2011}.  Assuming that the
interstellar matter is composed primarily of hydrogens and heliums,
the deceleration radius, $R_{\rm dec}$, for spherical homologous
ejecta is calculated~\cite{NP2011,TKI,Hotoke16r}, and then the
deceleration time defined by $R_{\rm dec}/\bar v_{\rm ej}$ is given by
\beqn
t_{\rm dec} &\approx& 45\,{\rm yrs}\,
\biggl({E_0 \over 6 \times 10^{50}\,{\rm erg}}\biggr)^{1/3}
\biggl({n_0 \over 0.01\,{\rm cm}^{-3}}\biggr)^{-1/3} \nonumber \\
&&~~~~~~~~~~~\times
\biggl({\bar v_{\rm ej} \over 0.15c}\biggr)^{-5/3}, \label{eq32}
\eeqn
where $n_0$ is the number density of the interstellar matter (ISM) and 
$E_0=M_{\rm ej}\bar v_{\rm ej}^2/2$ is the total kinetic energy of
the ejecta.  For $M_{\rm ej}=0.03M_\odot$ and $\bar v_{\rm ej}=0.15c$, 
\beqn
E_0=6 \times 10^{50}\,{\rm erg}
\left({M_{\rm ej} \over 0.03M_\odot}\right)
\left({\bar v_{\rm ej} \over 0.15c}\right)^2.
\eeqn
Thus, the radio flare associated with the ejecta is expected to reach
the peak approximately at $\sim 45(n_0/0.01\,{\rm
  cm}^{-3})^{-1/3}$\,yrs after the merger. 

For the typical value of the ejecta velocity $\bar v_{\rm ej} \sim
0.15c$, the peak flux for the observed frequency is obtained at the
deceleration time described in Eq.~(\ref{eq32}).  The peak flux for a
given observed radio-band frequency $\nu_\mathrm{obs}$ is estimated
as~\cite{NP2011}
\beqn
F_\nu &\approx& 190\,\mu\mathrm{Jy} 
\left( {E_0 \over 10^{51}\,\mathrm{erg}} \right) 
\left( {n_0 \over 0.01\,\mathrm{cm}^{-3}} \right)^{(p+1)/4} 
\nonumber \\
&&~~ \times
\left( {\varep_e \over 0.1} \right)^{p-1} 
\left( {\varep_B \over 0.1} \right)^{(p+1)/4} 
\left( {\bar v_{\rm ej} \over 0.15c} \right)^{(5p-7)/2}
\nonumber \\
&&~~ \times
\left( {D \over 40\,\mathrm{Mpc}} \right)^{-2} 
\left( {\nu_\mathrm{obs} \over 1.4\,\mathrm{GHz}} \right)^{-(p-1)/2},~~~
\label{eq33}
\eeqn 
where we assumed the power-law distribution of the electron's Lorentz
factor with the power-law index $p=2.5$ for deriving the specific
value, and $\varep_e$ and $\varep_B$ denote the energy fractions of
accelerated electrons and magnetic-field in the shock, respectively.
Equation~(\ref{eq33}) is applicable as long as the observed frequency
is higher than the typical synchrotron and self-absorption frequency
at the deceleration time, $t_{\rm dec}$.  Equation~(\ref{eq33}) shows
that the peak flux is high enough for the radio telescope to detect
the signal, even if $n_0$ is not very high as $\sim 0.01\,{\rm
  cm^{-3}}$~\cite{Alex17}.
  
We note that the dynamical ejecta could have a velocity distribution
in a broad range up to $\approx 0.8c$ and this fast component with the
ejecta velocity $v_{\rm ej}\agt 0.5c$ could have appreciable mass of
$\sim 10^{-5}$--$10^{-4}M_\odot$ (i.e., the kinetic energy is $\sim 2
\times 10^{48}$--$2\times 10^{49}\,{\rm erg}$)~\cite{Hotoke13a}. Its
deceleration time is much shorter as $\sim$ (1--2)$(n_0/0.01\,{\rm
  cm}^{-3})^{-1/3}$\,yr. Therefore the radio light curve arising from
the dynamical ejecta is likely to have a broad peak~\cite{Hotoke16r}.
The peak flux arising from this early fast component is estimated as
\beqn
F_\nu &\approx& 30\,\mu\mathrm{Jy} 
\left( {E_{0,{\rm f}} \over 5\times 10^{48}\,\mathrm{erg}} \right) 
\left( {n_0 \over 0.01\,\mathrm{cm}^{-3}} \right)^{(p+1)/4} 
\nonumber \\
&&~~ \times
\left( {\varep_e \over 0.1} \right)^{p-1} 
\left( {\varep_B \over 0.1} \right)^{(p+1)/4} 
\left( {v_{\rm ej,f} \over 0.5c} \right)^{(5p-7)/2}
\nonumber \\
&&~~ \times
\left( {D \over 40\,\mathrm{Mpc}} \right)^{-2} 
\left( {\nu_\mathrm{obs} \over 1.4\,\mathrm{GHz}} \right)^{-(p-1)/2},~~~
\label{eq34}
\eeqn 
where $E_{0,{\rm f}}$ and $v_{\rm ej,f}$ denote the kinetic energy and
velocity for the fast component. The peak time of the flux arising
from this component is earlier than the deceleration time estimated by
Eq.~(\ref{eq32}) because the difference between the observer time and
the time in the ejecta frame is significant for the ejecta with such
high velocities.  The rise rate of the radio flux is shallower than
$\propto t^3$ due to the contribution of the shells with different
velocities~\cite{Hotoke16r}.  Depending on the ISM density and the
velocity distribution, the radio signal can be detected, even far
before the peak time described in Eq.~(\ref{eq32}) for GW170817.  The
detection of the radio flare at early times is quite important for
proving the dynamical ejecta of the merger. Since the mass and
velocity for the early component of the dynamical ejecta depend
strongly on the neutron-star EOS (faster material is ejected for more
compact neutron stars), the luminosity as a function of time will also
carry information for the EOS.

The detection of the early radio signals reported by
Refs.~\cite{Hallinan17,Alex17} is not likely to be associated with the
sub-relativistic mass ejection with $\bar v_{\rm ej}\sim0.2c$ unless
$n_0 \gg 0.01\,{\rm cm}^{-3}$.  Note that the density inferred from
the limit of the HI observation is $<10^{-2}\,{\rm cm^{-3}}$.
Therefore this early radio signal is likely to be associated with some
relativistic mass ejection with $\bar v_{\rm ej} \sim
c$~\cite{Hallinan17,Alex17}.  We speculate that the radio flare
associated with the mass ejection of fast motion with $v_{\rm ej} \agt
0.5c$ will be detected by subsequent observations in a few years.

\subsection{Perspective for possible future events}

If the inclination angle of the rotational axis of the binary orbital
motion with respect to our line of sight were close to $\sim
90^\circ$, the observational properties of the electromagnetic
counterparts of GW170817 would be significantly different from those
of the electromagnetic counterparts of this event, because lanthanide
elements are likely to be present along our line of sight.  If so, the
electromagnetic counterpart would be much less luminous and the time
to reach the peak luminosity would be delayed because we could only
observe the ejecta of high opacity $\kappa \sim 10\,{\rm cm^2/g}$. For
such events, the ratio of the effective distance, $D_{\rm eff}$, to
the luminosity distance, $D$, to the source should be larger than
$\sim 1.5$, and hence, the SNR for the gravitational-wave observation
would be smaller than that for GW170817 for a given value of $D$;
i.e., the observation would be less frequent. However, in the future
for which the sensitivity of the gravitational-wave detectors is
improved significantly, such edge-on events will be detected by the
gravitational-wave detectors, and in such a forthcoming case, a
macronova/kilonova, for which the feature is different from that of
GW170817 even for the same mass of binary components, will be
observed~(see also Ref.~\cite{Brian17}).

As discussed in Sec.~\ref{sec3}, the remnant for the merger in the
GW170817 event would be a long-lived massive neutron star surrounded by
a torus. However, this may be the case only for $m \alt 2.8M_\odot$ even
for the stiff EOS.  For events with $m \agt 2.8M_\odot$, the remnant may
be a black hole surrounded by a torus.  As described in
Sec.~\ref{sec3.2}, for this case, the ejecta could always be composed of
neutron-rich matter with $Y_e \alt 0.2$. Then, the opacity of the ejecta
should be high, $\kappa \sim 10\,{\rm cm^2/g}$, and hence, the peak time
and peak luminosity of the electromagnetic counterparts could be
significantly different from those for GW170817. When such a
macronova/kilonova is discovered associated with a gravitational-wave
detection for the merger of binary neutron stars, the results for the
electromagnetic counterparts together with the total mass of the binary
system will be used to constrain the neutron-star EOS.

As mentioned in Sec. IV B, the electromagnetic observation for
GW170817 suggests that a remnant massive neutron star collapses to a
black hole before a substantial fraction of its rotational kinetic
energy is dissipated through the magnetic dipole radiation. However,
for an event in which the total mass of a system (and as a result,
mass of the remnant massive neutron star) is smaller than that for
GW170817, the remnant massive neutron star may survive for a longer
time scale of $\agt 100$\,s. In such a case, a large fraction of its
rotational kinetic energy may be released by the magnetic dipole
radiation, leading to the acceleration of the ejecta to a relativistic
speed.  As a result, we expect a strong synchrotron radiation, that
peaks at the radio bands, arising from the forward shock~\cite{MB2014,
  Horesh16} and a magnetar-wind nebula producing the optical and X-ray
emission~\cite{Metzger2014}.  In the future observation, this type of
the event could be found for the merger of low-mass binary neutron
stars. The observation of such an event will be also used to constrain
the neutron-star EOS because we can obtain a lower bound for the
maximum mass of rapidly rotating neutron stars. Thus, the future
observations for a variety of the binary neutron star mergers will
significantly narrow down the possibility for the neutron-star EOS.

\section{Summary}\label{sec5}

In this paper, we attempt to interpret the observational results for
the electromagnetic counterparts of GW170817 using the results of our
numerical-relativity simulations performed so far.  The characteristic
features for the electromagnetic counterparts are their early peak
time and high luminosity in the optical to IR bands.  The
numerical-relativity results indicate that a long-lived massive
neutron star surrounded by a torus is a favored remnant for
interpreting this event, because only in the presence of such a strong
neutrino emitter, the major ejecta component of sufficiently large
mass of $\sim 0.03M_\odot$ can have a sufficiently high electron
fraction of $Y_e \agt 0.25$, avoiding the enhancement of the ejecta
opacity.  The long-lived massive neutron star also plays a role for
ejecting appreciable amount of material of fast motion.  For getting
such merger remnants, an EOS with a reasonably high value of $M_{\rm
  max}$ is required. No detection of relativistic optical counterpart
suggests a value of $M_{\rm max}$ approximately to be
2.15--$2.25M_\odot$.

As discussed in Sec.~\ref{sec3.2}, if the remnant of the merger is a
black hole surrounded by a torus, we may not have a strong emitter of
neutrinos. If so, it would not reproduce the electromagnetic
observational results of GW170817. However, it is not currently clear
whether the torus is really a weak emitter of neutrinos or not. Some
numerical experiments suggest that the torus surrounding a spinning
black hole could be a strong emitter of neutrinos with the luminosity
appreciably larger than $10^{52}\,{\rm erg/s}$, if the torus mass is
sufficiently large. To date, we have not had detailed general
relativistic radiation hydrodynamics simulations for such systems.  We
plan to perform simulations for this system in the near future.

Also, we plan to perform a variety of simulations fixing the chirp
mass of ${\cal M} \approx 1.19M_\odot$ but employing new EOSs like in
Ref.~\cite{Togashi} and changing mass ratio, $q$, for a wide
range. Our present study indicates that the merger remnant for the
GW170817 event should be a strong emitter of neutrinos like a
long-lived massive neutron star.  To form a massive neutron star from
binaries of total mass $\agt 2.73M_\odot$, a stiff EOS is
necessary. This suggests that soft EOSs like the SFHo EOS may be
excluded.  Generally speaking, EOSs that produce a large-radius neutron
star are suitable for forming a long-lived massive neutron star as the
merger remnant.  However, even in the case that the typical radius is
not very large (e.g., $\sim 11$--$12$\,km) as suggested by the
analysis of the binary tidal deformability to GW170817, if the
maximum-allowed mass for cold spherical neutron stars is appreciably
lager than $2M_\odot$ (say $2.2M_\odot$~\cite{Togashi}), a long-lived
massive neutron star is likely to be the typical merger remnant. For
exploring this possibility, we need more simulations employing a
variety of neutron-star EOSs.

\begin{acknowledgments}

We thank Jim Lattimer, Brian Metzger, and Tsvi Piran for useful
discussions during a long-term workshop ``Electromagnetic Signatures
of $r$-process Nucleosynthesis in Neutron Star Binary Mergers'',
INT\,17-2b in Seattle. We also thank K. Ioka, K. Kashiyama, and
K. Kawaguchi for helpful discussions.  Numerical computation was
performed on K computer at AICS (project numbers hp160211 and
hp170230), on Cray XC30 at cfca of National Astronomical Observatory
of Japan, FX10 at Information Technology Center of the University of
Tokyo, HOKUSAI FX100 at RIKEN, and on Cray XC40 at Yukawa Institute
for Theoretical Physics, Kyoto University.  This work was supported by
Grant-in-Aid for Scientific Research (16H02183, JP16H06342,
JP17H01131, 17H06361, 15K05077) of JSPS and by a post-K computer
project (Priority issue No.~9) of Japanese MEXT. KH is supported by
Flatiron fellowship at the Simons Foundation and Lyman Spitzer Jr. 
Fellowship.

\end{acknowledgments}

\appendix

\section{Note on dynamical ejecta mass}

\begin{table*}[t]
\centering
\caption{\label{tab4} Comparison of the ejecta mass for a model with
  the SFHo EOS and $m_1=m_2=1.35M_\odot$.  Listed are locations of the
  outer boundary along each axis, $L$, the minimum grid spacing, $\varDelta x$,
  floor density, presence or absence of neutrino heating
  (irradiation), and dynamical ejecta mass.  The values in the
  parenthesis for the row of $L$ denote the location where the
  information of the ejecta is extracted.  ``No description'' means
  that no information is written on the corresponding point.  In the
  unpublished work by Shibata, Hotokezaka, Kyutoku, and Sekiguchi,
  they constructed a piecewise polytropic EOS for the SFHo EOS and
  performed a purely hydrodynamics simulation as done in
  Ref.~\cite{Hotoke13a}.  }
\begin{tabular}{lccccc}
\hline\hline
~Groups~ & ~$L$~(km)~ & ~$\varDelta x$~(m)~ & ~Floor density (${\rm g/cm^3}$)~ 
& ~Neutrino heating~ & ~Ejecta mass ($M_\odot$)~ \\
\hline\hline
Sekiguchi et al.~\cite{sekig15} & ~$10944$~ & 150 & $1.6 \times 10^4$ & Yes & 
$1.1 \times 10^{-2}$ \\
Sekiguchi et al.~\cite{sekig15} & ~$10240$~ & 250 & $1.6 \times 10^4$ & Yes &
$1.3 \times 10^{-2}$ \\
Sekiguchi et al.~\cite{sekig15} & ~$10240$~ & 250 & $1.6 \times 10^4$ & No &
$1.0 \times 10^{-2}$ \\
Palenzuela et al.~\cite{Palenzuela2015} & $750$ & 230 & $6 \times 10^5$ & No & 
$3.2 \times 10^{-3}$ \\
Bovard et el.~\cite{Bovard} & 760 (300) & 215 & no description & No & 
$3.5 \times 10^{-3}$ \\
Radice et al.~\cite{radice2017} & 1512 (433) & 185 & no description & No 
& $3.5 \times 10^{-3}$ \\
\hline 
Shibata et al. (unpublished) &  2858 & 186 & $8.2 \times 10^3$ & No &  
$1.1 \times 10^{-2}$ \\
\hline\hline
\end{tabular}
\end{table*}

Because our referee suggests us to compare the backgrounds in
general-relativistic radiation hydrodynamics simulations for the study
of dynamical ejecta by different groups, we list several quantities
for a specific model in Table~\ref{tab4}.  In this model, the SFHo EOS
is employed and each mass of the binary is $m_1=m_2=1.35M_\odot$. All
the groups employed a mesh-refinement algorithm, but the location of
the outer boundary along each axis and minimum grid spacing are
different among different groups, in particular between ours and other
groups. The floor density has to be put in a dilute-density or vacuum
region outside the neutron stars and merger remnant when using the
conservative form of hydrodynamics in numerical simulations. Its
choice is one of the crucial artificial points for accurately
exploring the mass ejection during the merger process and is also
likely to be different among four groups.  Finally, most groups except
for ours performed simulations simply using a leakage scheme without
taking into account neutrino heating. The neutrino heating is crucial
for exploring the values of the electron fraction and nucleosynthesis
in the ejecta.  Also, difference in the treatments of neutrino physics
affects the properties of the ejecta.  According to Foucart et
al.~\cite{Foucart2016a}, the ejecta mass and the electron fraction of the
polar ejecta could be changed by $\sim 20\%$ and $\gtrsim 50\%$ due to
the treatment of neutrinos in the energy-integrated radiation
transport scheme (specifically the definition of the neutrino energy
could affect the results).  It should be also mentioned that the
neutrino leakage scheme employed by us is different from other groups, 
all of which use a similar scheme.  Currently it is difficult to
quantify the uncertainty caused by the neutrino treatment because of
the lack of reliable results based on more physical neutrino transport
schemes.

Table~\ref{tab4} shows that the setup for the simulations is
significantly different among four groups. The three groups except for
ours located the outer boundaries at a region fairly close to the
center, and estimated the ejecta mass essentially in the near zone.
If a part of the ejecta component is produced by getting energy in a
far region, e.g., by angular momentum transport due to tidal torque
exerted by the central object and neutrino heating, the ejecta mass
could be underestimated for the simulations with a small computational
region (see also a discussion of Ref.~\cite{Bovard}).  It should be
also noted that these groups estimated the ejecta mass before the
spacetime in a corresponding region relaxes to a stationary state
because the typical velocity of the dynamical ejecta is $v_{\rm ej}
\sim 0.2c$--$0.25c$, i.e., $\approx 60$--75\,km/ms, and the ejecta
goes through the outer boundaries or the surface of the flux integral
at $\alt 10$\,ms after the onset of merger.  We here note that for
estimating the ejecta mass, we usually employ the condition of $u_t <
-1$ or $h u_t < -1$ where $u_t$ is the lower time component of the
four velocity and $h$ is the specific enthalpy. When we employ this
method, the spacetime has to be stationary (a time-like Killing vector
has to be present) but it is not clear whether the stationarity is
well established at $\alt 10$\,ms after the onset of merger. 

On the other hand, we prepared a wider computational domain and
calculated the ejecta mass by the direct volume integral at late time,
i.e., at 20--30\,ms after the onset of merger~\cite{sekig15}.  We note
that for employing this method, the computational region is wide 
enough ($L \agt v_{\rm ej}\times 30\,{\rm ms} \sim 2000$\,km) and
floor density is small enough (total floor mass is much smaller than
the ejecya mass for $L \alt 2000$\,km).  We then confirmed that the 
ejecta mass depends weakly on the time of the estimation. We caution
that the ejecta mass depends strongly on the time of estimation, if we
estimated it at $\alt 10$\,ms after the onset of merger~(see Fig.~1 of
Ref.~\cite{sekig15}): The estimation at earlier time results in
smaller ejecta mass of $<0.01M_\odot$.

Our numerical results indicate that the finer grid resolution results
in smaller ejecta mass (see Table~\ref{tab4}). Thus, in reality, the
ejecta mass may be slightly smaller than $0.011M_\odot$ in our
radiation-hydrodynamics implementation.  Our numerical results also
show that in the presence of neutrino heating, the ejecta mass is
increased by 30\% (see Table~\ref{tab4}). Thus, if other groups take
into account this effect, their estimation for the dynamical ejecta
mass may be increased.

To cross-check our result on the ejecta mass, we also performed a purely
hydrodynamics simulation employing a piecewise polytropic EOS model for
the SFHo EOS as done in Ref.~\cite{Hotoke13a} (see the last column of
Table~\ref{tab4}). For this case, the ejecta mass agrees broadly with
that in Ref.~\cite{sekig15}. This suggests that radiation hydrodynamics
effects would play a subdominant role for determining the ejecta mass
if it is as large as $\sim 0.01 M_\odot$. However, these effects
are likely to become appreciable when the mass of dynamical ejecta is
small.

As found from the above discussion, obviously, comparison of the
numerical results by different groups cannot be currently done in a
well-defined manner because the computational setup is significantly
different among them. In the future, we need comparison employing the
same computational region, grid spacing, and floor density with the
same neutrino physics.  In the absence of such comparison works, it is
safe to keep in mind that there is an uncertainty of a factor $\-sim 2$
in the estimation of the dynamical ejecta mass.



\end{document}